\documentclass[11pt]{article}
\def\half{\mbox{\small $\frac{1}{2}$}}

\def\vec#1{\ifmmode
\mathchoice{\mbox{\boldmath$\displaystyle\bf#1$}}
{\mbox{\boldmath$\textstyle\bf#1$}}
{\mbox{\boldmath$\scriptstyle\bf#1$}}
{\mbox{\boldmath$\scriptscriptstyle\bf#1$}}\else
{\mbox{\boldmath$\bf#1$}}\fi}

\begin{document}

\begin{flushleft}
\end{flushleft}

\vspace{2 cm}

\begin{center}
{\Large Asymptotic formulae for likelihood-based tests of new physics}
\end{center}

\vspace{3 cm}

\begin{center}
Glen Cowan$^1$, Kyle Cranmer$^2$,
Eilam Gross$^3$, Ofer Vitells$^3$
\end{center}

\vspace{0.5 cm}

\noindent
$^1$  Physics Department, Royal Holloway, University of London,
Egham, TW20 0EX, U.K. \\
$^2$   Physics Department, New York University, New York, NY 10003,
U.S.A. \\
$^3$  Weizmann Institute of Science, Rehovot 76100, Israel

\vspace{3 cm}


\begin{abstract}
We describe likelihood-based statistical tests for use in high energy
physics for the discovery of new phenomena and for construction
of confidence intervals on model parameters.  We focus on the
properties of the test procedures that allow one to account for
systematic uncertainties.  Explicit formulae for the asymptotic
distributions of test statistics are derived using results of Wilks
and Wald.  We motivate and justify the use of a representative data
set, called the ``Asimov data set'', which provides a simple method to
obtain the median experimental sensitivity of a search or measurement
as well as fluctuations about this expectation.
\end{abstract}

\vspace{1 cm}

\clearpage

\section{Introduction}
\label{sec:intro}

In particle physics experiments one often searches for processes that
have been predicted but not yet seen, such as production of a Higgs
boson.  The statistical significance of an observed signal can be
quantified by means of a $p$-value or its equivalent Gaussian
significance (discussed below).  It is useful to characterize the
sensitivity of an experiment by reporting the expected (e.g., mean or
median) significance that one would obtain for a variety of signal
hypotheses.

Finding both the significance for a specific data set and the expected
significance can involve Monte Carlo calculations that are
computationally expensive.  In this paper we investigate approximate
methods based on results due to Wilks \cite{Wilks} and Wald
\cite{Wald} by which one can obtain both the significance for given
data as well as the full sampling distribution of the significance
under the hypothesis of different signal models, all without recourse
to Monte Carlo.  In this way one can find, for example, the median
significance and also a measure of how much one would expect this to
vary as a result of statistical fluctuations in the data.

A useful element of the method involves estimation of the median
significance by replacing the ensemble of simulated data sets by a
single representative one, referred to here as the ``Asimov'' data
set.%
\footnote{The name of the Asimov data set is inspired by the short
story {\it Franchise}, by Isaac Asimov \cite{Asimov}. In it, elections
are held by selecting the single most representative voter to replace
the entire electorate.}  In the past, this method has been used and
justified intuitively (e.g., \cite{quast,CSC}).  Here we provide a
formal mathematical justification for the method, explore its
limitations, and point out several additional aspects of its use.

The present paper extends what was shown in Ref.~\cite{CSC} by giving
more accurate formulas for exclusion significance and also by
providing a quantitative measure of the statistical fluctuations in
discovery significance and exclusion limits.  For completeness some of
the background material from \cite{CSC} is summarized here.

In Sec.~\ref{sec:formalism} the formalism of a search as a
statistical test is outlined and the concepts of statistical
significance and sensitivity are given precise definitions.
Several test statistics based on the profile likelihood
ratio are defined.

In Sec.~\ref{sec:qdist}, we use the approximations due to Wilks and
Wald to find the sampling distributions of the test statistics
and from these find $p$-values and related quantities for
a given data sample.
In Sec.~\ref{sec:sensitivity} we discuss how to determine the
median significance that one would obtain for an assumed
signal strength.
Several example applications are shown in Sec.~\ref{sec:examples}, and
numerical implementation of the methods in the RooStats package is
described in Sec.~\ref{sec:roostats}.  Conclusions are given in
Sec.~\ref{sec:conclusions}.

\section{Formalism of a search as a statistical test}
\label{sec:formalism}

In this section we outline the general procedure used to search for a
new phenomenon in the context of a frequentist statistical test.  For
purposes of discovering a new signal process, one defines the null
hypothesis, $H_0$, as describing only known processes, here designated
as background.  This is to be tested against the alternative $H_1$,
which includes both background as well as the sought after signal.
When setting limits, the model with signal plus background plays the
role of $H_0$, which is tested against the background-only hypothesis,
$H_1$.

To summarize the outcome of such a search one quantifies the level of
agreement of the observed data with a given hypothesis $H$ by
computing a $p$-value, i.e., a probability, under assumption of $H$,
of finding data of equal or greater incompatibility with the
predictions of $H$.  The measure of incompatibility can be based, for
example, on the number of events found in designated regions of
certain distributions or on the corresponding likelihood ratio for
signal and background. One can regard the hypothesis as excluded if
its $p$-value is observed below a specified threshold.

In particle physics one usually converts the $p$-value into an
equivalent significance, $Z$, defined such that a Gaussian distributed
variable found $Z$ standard deviations above\footnote{Some authors, e.g.,
Ref.~\cite{lepcombo}, have defined this relation using a two-sided
fluctuation of a Gaussian variable, with a $5 \sigma$ significance
corresponding to $p = 5.7 \times 10^{-7}$.  We take the one-sided
definition above as this gives $Z=0$ for $p=0.5$.
}
its mean has an upper-tail probability equal to $p$.  That is,

\begin{equation}
\label{eq:significance}
Z = \Phi^{-1}(1-p) \,,
\end{equation}

\noindent where $\Phi^{-1}$ is the quantile (inverse of the cumulative
distribution) of the standard Gaussian.  For a signal process such as
the Higgs boson, the particle physics community has tended to regard
rejection of the background hypothesis with a significance of at least
$Z=5$ as an appropriate level to constitute a discovery.  This
corresponds to $p = 2.87 \times 10^{-7}$.  For purposes of
excluding a signal hypothesis, a threshold $p$-value of 0.05 (i.e.,
95\% confidence level) is often used, which corresponds to $Z = 1.64$.


It should be emphasized that in an actual scientific context,
rejecting the background-only hypothesis in a statistical sense is
only part of discovering a new phenomenon.  One's degree of belief
that a new process is present will depend in general on other factors
as well, such as the plausibility of the new signal hypothesis and the
degree to which it can describe the data.  Here, however, we only
consider the task of determining the $p$-value of the background-only
hypothesis; if it is found below a specified threshold, we regard this
as ``discovery''.

It is often useful to quantify the sensitivity of an experiment by
reporting the expected significance one would obtain with a given
measurement under the assumption of various hypotheses.  For example,
the sensitivity to discovery of a given signal process $H_1$ could be
characterized by the expectation value, under the assumption of $H_1$,
of the value of $Z$ obtained from a test of $H_0$.  This would not be
the same as the $Z$ obtained using Eq.~(\ref{eq:significance}) with
the expectation of the $p$-value, however, because the relation
between $Z$ and $p$ is nonlinear.  The median $Z$ and $p$ will,
however, satisfy Eq.~(\ref{eq:significance}) because this is a
monotonic relation.  Therefore in the following we will take the term
`expected significance' always to refer to the median.




A widely used procedure to establish discovery (or exclusion) in
particle physics is based on a frequentist significance test using a
likelihood ratio as a test statistic.  In addition to parameters of
interest such as the rate (cross section) of the signal process, the
signal and background models will contain in general {\it nuisance
parameters} whose values are not taken as known {\it a priori} but
rather must be fitted from the data.

It is assumed that the parametric model is sufficiently flexible so
that for some value of the parameters it can be regarded as true.  The
additional flexibility introduced to parametrize systematic effects
results, as it should, in a loss in sensitivity.  To the degree
that the model is not able to reflect the truth accurately, an
additional systematic uncertainty will be present that is not
quantified by the statistical method presented here.



To illustrate the use of the profile likelihood ratio, consider an
experiment where for each selected event one measures the values of
certain kinematic variables, and thus the resulting data can be
represented as one or more histograms.  Using the method in an
unbinned analysis is a straightforward extension.

Suppose for each event in the signal sample one measures a variable
$x$ and uses these values to construct a histogram $\vec{n} = (n_1,
\ldots, n_N)$.  The expectation value of $n_i$ can be written

\begin{equation}
\label{eq:eni}
E[n_i] = \mu s_i + b_i \;,
\end{equation}

\noindent where the mean number of entries in the $i$th bin from signal
and background are

\begin{eqnarray}
\label{eq:si}
s_i = s_{\rm tot} \int_{{\rm bin} \, i} f_{s}(x; \vec{\theta}_{s}) \, dx \,,
\\*[0.3 cm]
\label{eq:bi}
b_i = b_{\rm tot} \int_{{\rm bin} \, i} f_{b}(x; \vec{\theta}_{b}) \, dx \,.
\end{eqnarray}

\noindent Here the parameter $\mu$ determines the strength of the
signal process, with $\mu=0$ corresponding to the background-only
hypothesis and $\mu=1$ being the nominal signal hypothesis.  The
functions $f_s(x;\vec{\theta}_s)$ and $f_b(x;\vec{\theta}_b)$ are the
probability density functions (pdfs) of the variable $x$ for signal
and background events, and $\vec{\theta}_s$ and $\vec{\theta}_b$
represent parameters that characterize the shapes of pdfs.  The
quantities $s_{\rm tot}$ and $b_{\rm tot}$ are the total mean numbers
of signal and background events, and the integrals in (\ref{eq:si})
and (\ref{eq:bi}) represent the probabilities for an event to be found
in bin $i$.  Below we will use $\vec{\theta} = (\vec{\theta}_s,
\vec{\theta}_b, b_{\rm tot})$ to denote all of the nuisance
parameters.  The signal normalization $s_{\rm tot}$ is not, however,
an adjustable parameter but rather is fixed to the value predicted by
the nominal signal model.

In addition to the measured histogram $\vec{n}$ one often makes
further subsidiary measurements that help constrain the nuisance
parameters.  For example, one may select a control sample where one
expects mainly background events and from them construct a histogram
of some chosen kinematic variable.  This then gives a set of values
$\vec{m} = (m_1, \ldots, m_M)$ for the number of entries in each of
the $M$ bins.  The expectation value of $m_i$ can be written

\begin{equation}
\label{eq:emi}
E[m_i] = u_i(\vec{\theta}) \;,
\end{equation}

\noindent where the $u_i$ are calculable quantities depending on the
parameters $\vec{\theta}$.  One often constructs this measurement so
as to provide information on the background normalization parameter
$b_{\rm tot}$ and also possibly on the signal and background shape
parameters.

The likelihood function is the product of Poisson probabilities for
all bins:

\begin{equation}
\label{eq:likelihood}
L(\mu, \vec{\theta}) =
\prod_{j=1}^N \frac{ (\mu s_{j} +
b_{j} )^{n_{j}} }{ n_{j}! }
e^{- (\mu s_{j} + b_{j}) }   \;\;
\prod_{k=1}^M \frac{ u_k^{m_{k}}} { m_{k}! } \,
e^{- u_k }  \;.
\end{equation}

To test a hypothesized value of $\mu$ we consider the
profile likelihood ratio

\begin{equation}
\label{eq:PLR}
\lambda(\mu) = \frac{ L(\mu,
\hat{\hat{\vec{\theta}}}) } {L(\hat{\mu}, \hat{\vec{\theta}}) } \;.
\end{equation}

\noindent Here $\hat{\hat{\vec{\theta}}}$ in the numerator denotes the
value of $\vec{\theta}$ that maximizes $L$ for the specified $\mu$,
i.e., it is the conditional maximum-likelihood (ML) estimator of
$\vec{\theta}$ (and thus is a function of $\mu$).  The denominator is
the maximized (unconditional) likelihood function, i.e., $\hat{\mu}$
and $\hat{\vec{\theta}}$ are their ML estimators. The presence of the
nuisance parameters broadens the profile likelihood as a function of
$\mu$ relative to what one would have if their values were fixed.
This reflects the loss of information about $\mu$ due to the
systematic uncertainties.

In many analyses, the contribution of the signal process to the mean
number of events is assumed to be non-negative.  This condition
effectively implies that any physical estimator for $\mu$ must be
non-negative.  Even if we regard this to be the case, however, it is
convenient to define an effective estimator $\hat{\mu}$ as the value
of $\mu$ that maximizes the likelihood, even this gives $\hat{\mu} <
0$ (but providing that the Poisson mean values, $\mu s_i + b_i$,
remain nonnegative).  This will allow us in Sec.~\ref{sec:wald} to
model $\hat{\mu}$ as a Gaussian distributed variable, and in this way
we can determine the distributions of the test statistics that we
consider.  Therefore in the following we will always regard
$\hat{\mu}$ as an effective estimator which is allowed to take on
negative values.

\subsection{Test statistic $t_{\mu} = - 2 \ln \lambda(\mu)$}
\label{sec:tmu}

From the definition of $\lambda(\mu)$ in Eq.~(\ref{eq:PLR}), one can
see that $0 \le \lambda \le 1$, with $\lambda$ near 1 implying good
agreement between the data and the hypothesized value of $\mu$.
Equivalently it is convenient to use the statistic

\begin{equation}
\label{eq:tmu}
t_{\mu} = -2 \ln \lambda(\mu)
\end{equation}

\noindent as the basis of a statistical test.  Higher values of
$t_{\mu}$ thus correspond to increasing incompatibility between the
data and $\mu$.

We may define a test of a hypothesized value of $\mu$ by using the
statistic $t_{\mu}$ directly as measure of discrepancy between the
data and the hypothesis, with higher values of $t_{\mu}$ correspond to
increasing disagreement.  To quantify the level of disagreement we
compute the $p$-value,

\begin{equation}
\label{eq:tmupval}
p_{\mu} = \int_{t_{\mu,{\rm obs}}}^{\infty} f(t_{\mu} | \mu ) \,
d t_{\mu} \;,
\end{equation}

\noindent where $t_{\mu,{\rm obs}}$ is the value of the statistic
$t_{\mu}$ observed from the data and $f(t_{\mu} | \mu )$ denotes the
pdf of $t_{\mu}$ under the assumption of the signal strength $\mu$.
Useful approximations for this and other related pdfs are given in
Sec.~\ref{sec:tmudist}.  The relation between the $p$-value and the
observed $t_{\mu}$ and also with the significance $Z$ are illustrated
in Fig.~\ref{fig:pval}.

\setlength{\unitlength}{1.0 cm}
\renewcommand{\baselinestretch}{0.9}
\begin{figure}[htbp]
\begin{picture}(10.0,4.5)
\put(1,0)
{\includegraphics{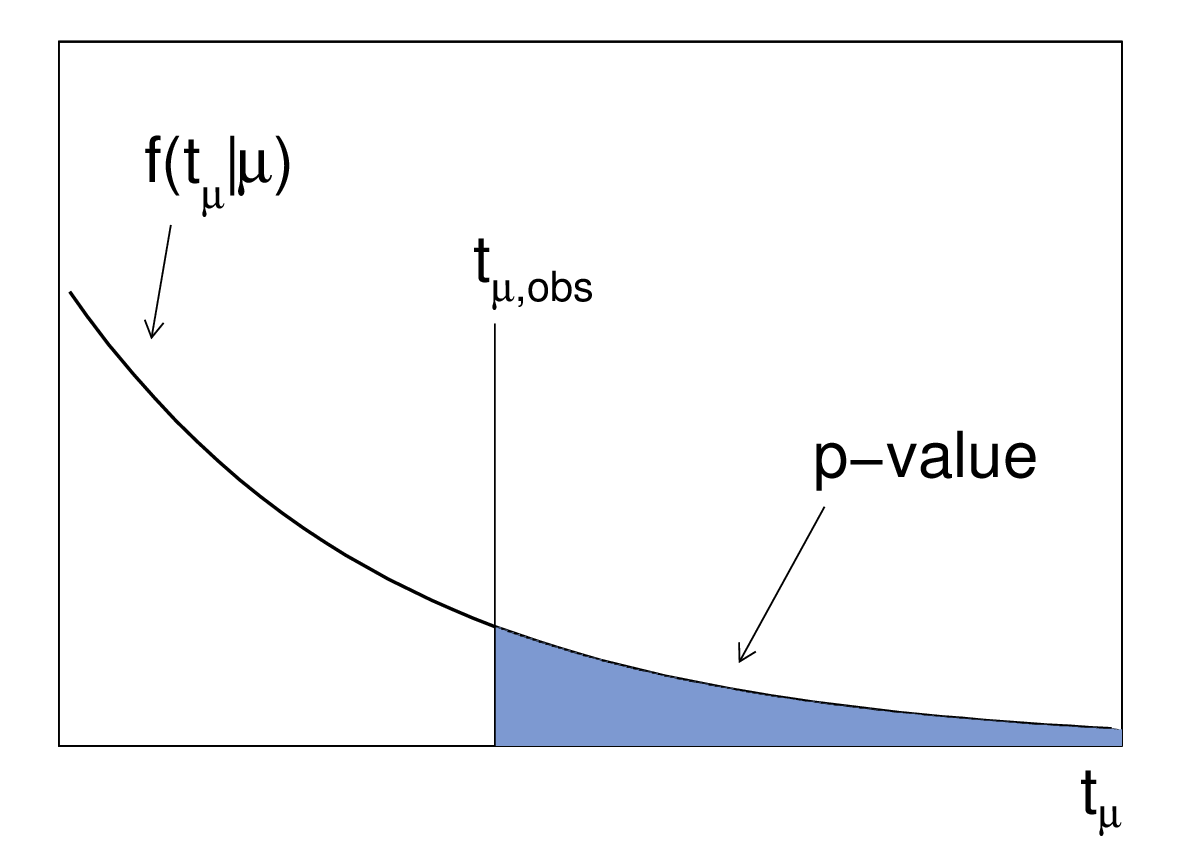}}
\put(8,0)
{\includegraphics{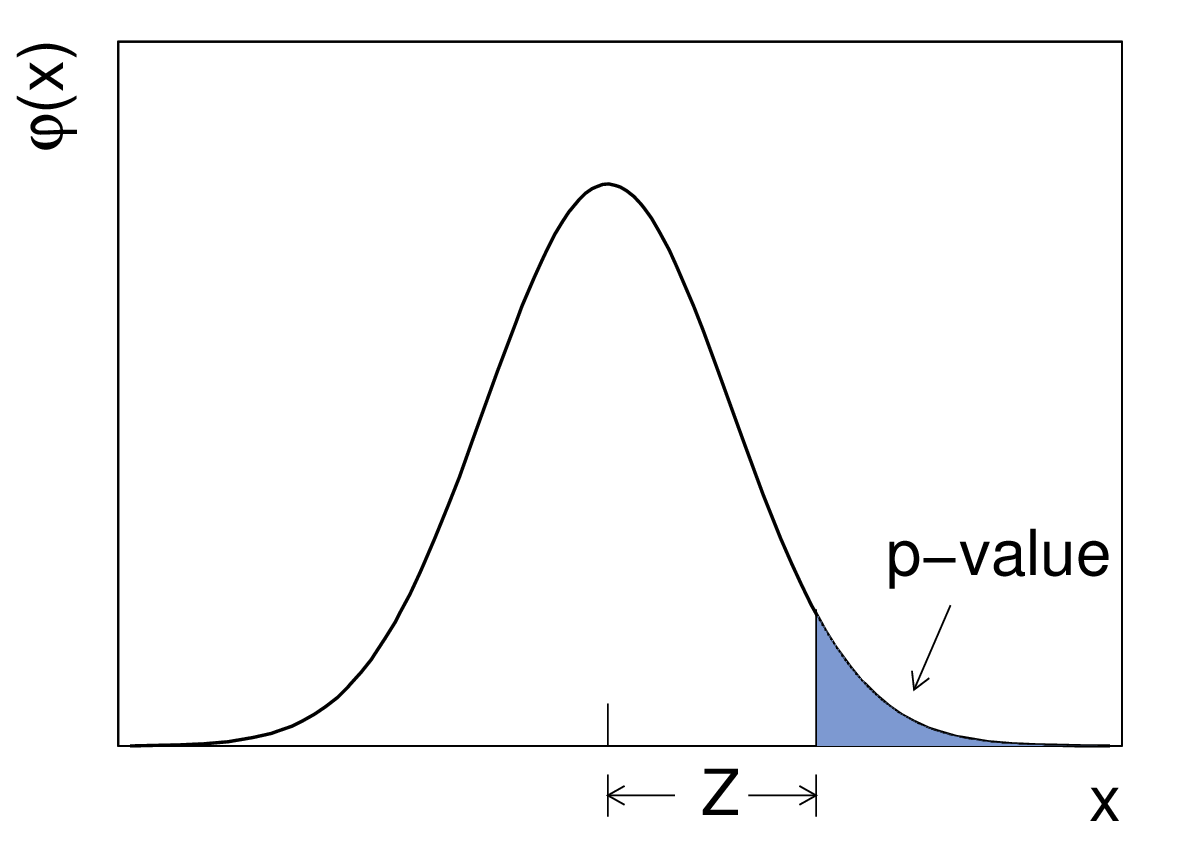}}
\put(0.3,3.8){(a)}
\put(14.5,3.8){(b)}
\end{picture}
\caption{\small (a) Illustration of the relation between the
$p$-value obtained from an observed value of the test statistic
$t_{\mu}$.   (b) The standard normal distribution $\varphi(x) = (1/\sqrt{2\pi})
\exp(-x^2/2)$ showing the relation between the significance $Z$ and
the $p$-value.}
\label{fig:pval}
\end{figure}
\renewcommand{\baselinestretch}{1}
\small\normalsize

When using the statistic $t_{\mu}$, a data set may result in a low
$p$-value in two distinct ways: the estimated signal strength
$\hat{\mu}$ may be found greater or less than the hypothesized value
$\mu$.  As a result, the set of $\mu$ values that are rejected because
their $p$-values are found below a specified threshold $\alpha$ may
lie to either side of those values not rejected, i.e., one may obtain
a two-sided confidence interval for $\mu$.

\subsection{Test statistic $\tilde{t}_{\mu}$ for $\mu \ge 0$}
\label{sec:tmutilde}

Often one assumes that the presence of a new signal can only
increase the mean event rate beyond what is expected from background
alone.  That is, the signal process necessarily has $\mu \ge 0$, and
to take this into account we define an alternative test statistic
below called $\tilde{t}_{\mu}$.


For a model where $\mu \ge 0$, if one finds data such that $\hat{\mu}
< 0$, then the best level of agreement between the data and any
physical value of $\mu$ occurs for $\mu = 0$.  We therefore define

\begin{equation}
\label{eq:lambdatilde}
\tilde{\lambda}({\mu}) =
\left\{ \! \! \begin{array}{ll}
               \frac{ L(\mu,
               \hat{\hat{\vec{\theta}}}(\mu)) }
               {L(\hat{\mu}, \hat{\vec{\theta}}) }
                 & \hat{\mu} \ge 0 , \\*[0.3 cm]
                \frac{ L(\mu,
               \hat{\hat{\vec{\theta}}}(\mu)) }
               {L(0, \hat{\hat{\vec{\theta}}}(0)) }
 & \hat{\mu} < 0 \;.
              \end{array}
       \right.
\end{equation}

\noindent Here $\hat{\hat{\vec{\theta}}}(0)$ and
$\hat{\hat{\vec{\theta}}}(\mu)$ refer to the conditional ML estimators
of $\vec{\theta}$ given a strength parameter of $0$ or $\mu$,
respectively.

The variable $\tilde{\lambda}(\mu)$ can be used instead
of $\lambda(\mu)$ in Eq.~(\ref{eq:tmu}) to obtain the corresponding
test statistic, which we denote $\tilde{t}_{\mu}$.  That is,

\begin{equation}
\label{eq:tmutilde}
\tilde{t}_{\mu} = - 2 \ln \tilde{\lambda}(\mu) =
\left\{ \! \! \begin{array}{ll}
               - 2 \ln \frac{L(\mu, \hat{\hat{\vec{\theta}}}(\mu))}
                {L(0, \hat{\hat{\theta}}(0))}
                & \quad \hat{\mu} < 0  \;, \\*[0.2 cm]
               -2 \ln \frac{L(\mu, \hat{\hat{\vec{\theta}}}(\mu))}
                {L(\hat{\mu}, \hat{\vec{\theta}})}
&  \quad \hat{\mu} \ge 0  \;.
              \end{array}
       \right.
\end{equation}

As was done with the statistic $t_{\mu}$, one can quantify the level
of disagreement between the data and the hypothesized value of $\mu$
with the $p$-value, just as in Eq.~(\ref{eq:tmupval}).  For this one
needs the distribution of $\tilde{t}_{\mu}$, an approximation of which
is given in Sec.~\ref{sec:tmutildedist}.

Also similar to the case of $t_{\mu}$, values of $\mu$ both above and
below $\hat{\mu}$ may be excluded by a given data set, i.e., one may
obtain either a one-sided or two-sided confidence interval for $\mu$.
For the case of no nuisance parameters, the test variable
$\tilde{t}_{\mu}$ is equivalent to what is used in constructing
confidence intervals according to the procedure of Feldman and Cousins
\cite{FC}.

\subsection{Test statistic $q_0$ for discovery of a positive signal}
\label{sec:discovery}

An important special case of the statistic $\tilde{t}_{\mu}$ described
above is used to test $\mu=0$ in a class of model where we assume $\mu
\ge 0$.  Rejecting the $\mu=0$ hypothesis effectively leads to the
discovery of a new signal.  For this important case we use the special
notation $q_0 = \tilde{t}_0$.  Using the definition
(\ref{eq:tmutilde}) with $\mu=0$ one finds

\begin{equation}
\label{eq:q0}
q_{0} =
\left\{ \! \! \begin{array}{ll}
               - 2 \ln \lambda(0)
               & \quad \hat{\mu} \ge 0 \;, \\*[0.3 cm]
               0 & \quad \hat{\mu} < 0  \;,
              \end{array}
       \right.
\end{equation}

\noindent where $\lambda(0)$ is the profile likelihood ratio for
$\mu=0$ as defined in Eq.~(\ref{eq:PLR}).

We may contrast this to the statistic $t_{0}$, i.e.,
Eq.~(\ref{eq:tmu}), used to test $\mu = 0$.  In that case one may
reject the $\mu=0$ hypothesis for either an upward or downward
fluctuation of the data.  This is appropriate if the presence of a new
phenomenon could lead to an increase or decrease in the number of
events found.  In an experiment looking for neutrino oscillations, for
example, the signal hypothesis may predict a greater or lower event
rate than the no-oscillation hypothesis.

When using $q_0$, however, we consider the data to show lack of
agreement with the background-only hypothesis only if $\hat{\mu} > 0$.
That is, a value of $\hat{\mu}$ much below zero may indeed constitute
evidence against the background-only model, but this type of
discrepancy does not show that the data contain signal events, but
rather points to some other systematic error.  For the present
discussion, however, we assume that the systematic uncertainties are
dealt with by the nuisance parameters $\vec{\theta}$.

If the data fluctuate such that one finds fewer events than even
predicted by background processes alone, then $\hat{\mu} < 0$ and one
has $q_0 = 0$.  As the event yield increases above the expected
background, i.e., for increasing $\hat{\mu}$, one finds increasingly
large values of $q_0$, corresponding to an increasing level of
incompatibility between the data and the $\mu=0$ hypothesis.

To quantify the level of disagreement between the data and the
hypothesis of $\mu = 0$ using the observed value of $q_0$ we compute
the $p$-value in the same manner as done with $t_{\mu}$, namely,

\begin{equation}
\label{eq:q0pval}
p_{0} = \int_{q_{0,{\rm obs}}}^{\infty} f(q_{0} | 0 ) \, d q_{0} \;.
\end{equation}

\noindent Here $f(q_{0} | 0 )$ denotes the pdf of the statistic $q_0$
under assumption of the background-only ($\mu=0$) hypothesis.  An
approximation for this and other related pdfs are given in
Sec.~\ref{sec:q0dist}.

\subsection{Test statistic $q_{\mu}$ for upper limits}
\label{sec:qmu}

For purposes of establishing an upper limit on the strength parameter
$\mu$, we consider two closely related test statistics.  First, we may
define

\begin{equation}
\label{eq:qmu}
q_{\mu} =
\left\{ \! \! \begin{array}{ll}
               - 2 \ln \lambda(\mu)  & \hat{\mu} \le \mu  \;, \\*[0.2 cm]
               0 & \hat{\mu} > \mu \;,
              \end{array}
       \right.
\end{equation}

\noindent where $\lambda(\mu)$ is the profile likelihood ratio as
defined in Eq.~(\ref{eq:PLR}).  The reason for setting $q_{\mu} = 0$
for $\hat{\mu} > \mu$ is that when setting an upper limit, one
would not regard data with $\hat{\mu} > \mu$ as representing less
compatibility with $\mu$ than the data obtained, and therefore this is
not taken as part of the rejection region of the test.  From the
definition of the test statistic one sees that higher values of
$q_{\mu}$ represent greater incompatibility between the data and the
hypothesized value of $\mu$.

One should note that $q_0$ is not simply a special case of $q_{\mu}$
with $\mu=0$, but rather has a different definition (see
Eqs.~(\ref{eq:q0}) and (\ref{eq:qmu})).  That is, $q_0$ is zero if the
data fluctuate downward ($\hat{\mu} < 0$), but $q_{\mu}$ is zero if
the data fluctuate upward ($\hat{\mu} > \mu$).  With that caveat in
mind, we will often refer in the following to $q_{\mu}$ with the idea
that this means either $q_0$ or $q_{\mu}$ as appropriate to the
context.

As with the case of discovery, one quantifies the level of agreement
between the data and hypothesized $\mu$ with $p$-value.  For, e.g., an
observed value $q_{\mu,{\rm obs}}$, one has

\begin{equation}
\label{eq:qmupval}
p_{\mu} = \int_{q_{\mu,{\rm obs}}}^{\infty} f(q_{\mu} | \mu) \, d q_{\mu} \;,
\end{equation}

\noindent which can be expressed as a significance using
Eq.~(\ref{eq:significance}).  Here $f(q_{\mu} | \mu)$ is the pdf of
$q_{\mu}$ assuming the hypothesis $\mu$.  In Sec.~\ref{sec:qmudist} we
provide useful approximations for this and other related pdfs.

\subsection{Alternative test statistic $\tilde{q}_{\mu}$ for upper limits}
\label{sec:qtildemu}

For the case where one considers models for which $\mu \ge 0$, the
variable $\tilde{\lambda}(\mu)$ can be used instead of $\lambda(\mu)$
in Eq.~(\ref{eq:qmu}) to obtain the corresponding test statistic,
which we denote $\tilde{q}_{\mu}$.  That is,

\begin{equation}
\label{eq:qmutilde}
\tilde{q}_{\mu} =
\left\{ \! \! \begin{array}{ll}
               - 2 \ln \tilde{\lambda}(\mu)  & \hat{\mu} \le \mu  \\*[0.2 cm]
               0 & \hat{\mu} > \mu
              \end{array}
       \right.
\quad = \quad \: \left\{ \! \! \begin{array}{lll}
               - 2 \ln \frac{L(\mu, \hat{\hat{\vec{\theta}}}(\mu))}
                {L(0, \hat{\hat{\theta}}(0))}
                & \hat{\mu} < 0 \;, \\*[0.2 cm]
               -2 \ln \frac{L(\mu, \hat{\hat{\vec{\theta}}}(\mu))}
                {L(\hat{\mu}, \hat{\vec{\theta}})}
&  0 \le \hat{\mu} \le \mu  \;, \\*[0.2 cm]
               0 & \hat{\mu} > \mu \;.
              \end{array}
       \right.
\end{equation}

\noindent We give an approximation for the pdf $f(\tilde{q}_{\mu} |
\mu^{\prime})$ in Sec.~\ref{sec:qtildemudist}.

In numerical examples we have found that the difference between the
tests based on $q_{\mu}$ (Eq.~(\ref{eq:qmu})) and $\tilde{q}_{\mu}$
usually to be negligible, but use of $q_{\mu}$ leads to important
simplifications.  Furthermore, in the context of the approximation
used in Sec.~\ref{sec:qdist}, the two statistics are equivalent.  That
is, assuming the approximations below, $q_{\mu}$ can be expressed as a
monotonic function of $\tilde{q}_{\mu}$ and thus they lead to the same
results.

\section{Approximate sampling distributions}
\label{sec:qdist}

In order to find the $p$-value of a hypothesis using
Eqs.~(\ref{eq:q0pval}) or (\ref{eq:qmupval}) we require the sampling
distribution for the test statistic being used.  In the case of
discovery we are testing the background-only hypothesis ($\mu=0$) and
therefore we need $f(q_0 | 0)$, where $q_0$ is defined by
Eq.~(\ref{eq:q0}).  When testing a nonzero value of $\mu$ for purposes
of finding an upper limit we need the distribution $f(q_{\mu} | \mu)$
where $q_{\mu}$ is defined by Eq.~(\ref{eq:qmu}), or alternatively we
require the pdf of the corresponding statistic $\tilde{q}_{\mu}$ as
defined by Eq.~(\ref{eq:qmutilde}).  In this notation the subscript of
$q$ refers to the hypothesis being tested, and the second argument in
$f(q_{\mu} | \mu)$ gives the value of $\mu$ assumed in the
distribution of the data.

We also need the distribution $f(q_{\mu} | \mu^{\prime})$ with $\mu
\ne \mu^{\prime}$ to find what significance to expect and how this is
distributed if the data correspond to a strength parameter different
from the one being tested.  For example, it is useful to characterize
the sensitivity of a planned experiment by quoting the median
significance, assuming data distributed according to a specified
signal model, with which one would expect to exclude the
background-only hypothesis.  For this one would need $f(q_0 |
\mu^{\prime} )$, usually with $\mu^{\prime} = 1$.  From this one can
find the median $q_0$, and thus the median discovery significance.
When considering upper limits, one would usually quote the value of
$\mu$ for which the median $p$-value is equal to 0.05, as this gives
the median upper limit on $\mu$ at 95\% confidence level.  In this
case one would need $f(q_{\mu} | 0)$ (or alternatively
$f(\tilde{q}_{\mu} | 0)$).

In Sec.~\ref{sec:wald} we present an approximation for the profile
likelihood ratio, valid in the large sample limit.  This allows one to
obtain approximations for all of the required distributions, which are
given in Sections~\ref{sec:tmudist} through \ref{sec:qmudist} The
approximations become exact in the large sample limit and are in fact
found to provide accurate results even for fairly small sample sizes.
For very small data samples one always has the possibility of using
Monte Carlo methods to determine the required distributions.

\subsection{Approximate distribution of the profile likelihood ratio}
\label{sec:wald}

Consider a test of the strength parameter $\mu$, which here can either
be zero (for discovery) or nonzero (for an upper limit), and suppose
the data are distributed according to a strength parameter
$\mu^{\prime}$.  The desired distribution $f(q_{\mu} | \mu^{\prime})$
can be found using a result due to Wald \cite{Wald}, who showed that
for the case of a single parameter of interest,

\begin{equation}
\label{eq:wald}
-2 \ln \lambda(\mu)
= \frac{(\mu - \hat{\mu})^2}{\sigma^2} + {\cal  O}(1/\sqrt{N}) \;.
\end{equation}

\noindent Here $\hat{\mu}$ follows a Gaussian distribution with a mean
$\mu^{\prime}$ and standard deviation $\sigma$, and $N$ represents the
data sample size.  The standard deviation $\sigma$ of $\hat{\mu}$ is
obtained from the covariance matrix of the estimators for all the
parameters, $V_{ij} = \mbox{cov}[\hat{\theta}_i, \hat{\theta}_j]$,
where here the $\theta_i$ represent both $\mu$ as well as the nuisance
parameters (e.g., take $\theta_0 = \mu$, so $\sigma^2 = V_{00}$).
In the large-sample limit, the bias of ML estimators in general tend to
zero, in which case we can write the inverse of the covariance matrix
as

\begin{equation}
\label{eq:vinvwald}
V^{-1}_{ij} = - E \left[ \frac{ \partial^2 \ln L }
{ \partial \theta_i \partial \theta_j } \right] \;,
\end{equation}

\noindent where the expectation value assumes a strength parameter
$\mu^{\prime}$.  The approximations presented here are valid to the
extent that the ${\cal O}(1/\sqrt{N})$ term can be neglected, and the
value of $\sigma$ can be estimated, e.g., using
Eq.~(\ref{eq:vinvwald}).
In Sec.~\ref{sec:asimov} we present an alternative way to estimate
$\sigma$ which lends itself more directly to determination of the
median significance.

%
%

If $\hat{\mu}$ is Gaussian distributed and we neglect the ${\cal
O}(1/\sqrt{N})$ term in Eq.~(\ref{eq:wald}), then one can show that
the statistic 
 $t_{\mu} = -2 \ln \lambda(\mu)$ follows a {\it noncentral
chi-square} distribution for one degree of freedom (see, e.g.,
\cite{ncc}),

\begin{equation}
\label{eq:ftmulambda}
f(t_{\mu};\Lambda) = \frac{1}{2 \sqrt{t_{\mu}}} \frac{1}{\sqrt{2 \pi}}
\left[ \exp \left( - \frac{1}{2}
\left( \sqrt{t_{\mu}} + \sqrt{\Lambda} \right)^2 \right) +
\exp \left( - \frac{1}{2} \left( \sqrt{t_{\mu}} - \sqrt{\Lambda} \right)^2
\right) \right] \;,
\end{equation}

\noindent where the noncentrality parameter $\Lambda$ is

\begin{equation}
\label{eq:noncentrality}
\Lambda = \frac{(\mu - \mu^{\prime})^2}{\sigma^2} \;.
\end{equation}

\noindent For the special case $\mu^{\prime} = \mu$ one has $\Lambda =
0$ and 
$-2 \ln\lambda(\mu)$ approaches a chi-square distribution
for one degree of freedom, a result shown earlier by Wilks
\cite{Wilks}.

The results of Wilks and Wald generalize to more than one parameter of
interest.  If the parameters of interest can be explicitly identified
with a subset of the parameters $\vec{\theta}_r = (\theta_1, \dots,
\theta_r$), then the distribution of $-2\ln\lambda(\vec{\theta}_r)$
follows a noncentral chi-square distribution for $r$-degrees of
freedom with noncentrality parameter

\begin{equation}
\label{eq:noncentralityND}
\Lambda_r   = \sum_{i,j=1}^{r} (\theta_i - \theta_i') \,\tilde{V}_{ij}^{-1}
\,(\theta_j - \theta_j') \;,
\end{equation}

\noindent where $\tilde{V}_{ij}^{-1}$ is the inverse of the submatrix
one obtains from restricting the full covariance matrix to the
parameters of interest.  The full covariance matrix is given from
inverting Eq.~(\ref{eq:vinvwald}), and we show an efficient way to
calculate it in Sec.~\ref{sec:asimov}.


%
%

\subsection{The Asimov data set and the variance of $\hat{\mu}$}
\label{sec:asimov}

Some of the formulae given require the standard deviation $\sigma$ of
$\hat{\mu}$, which is assumed to follow a Gaussian distribution with a
mean of $\mu^{\prime}$.  Below we show two ways of estimating
$\sigma$, both of which are closely related to a special, artificial
data set that we call the ``Asimov data set''.

We define the Asimov data set such that when one uses it to evaluate
the estimators for all parameters, one obtains the true parameter
values.  Consider the likelihood function for the generic analysis
given by Eq.~(\ref{eq:likelihood}).  To simplify the notation in this
section we define

\begin{equation}
\label{eq:nui}
\nu_i = \mu^{\prime} s_i + b_i \;.
\end{equation}

\noindent Further let $\theta_0 = \mu$ represent the strength
parameter, so that here $\theta_i$ can stand for any of the
parameters.  The ML estimators for the parameters can be found by
setting the derivatives of $\ln L$ with respect to all of the
parameters equal to zero:

\begin{equation}
\label{eq:dlnldtheta}
\frac{ \partial \ln L}{\partial \theta_j}
= \sum_{i=1}^N \left( \frac{n_i}{\nu_i} - 1 \right)
\frac{ \partial \nu_i}{\partial \theta_j} +
\sum_{i=1}^M \left( \frac{m_i}{u_i} - 1 \right)
\frac{ \partial u_i}{\partial \theta_j} = 0 \;.
\end{equation}

\noindent This condition holds if the Asimov data, $n_{i,{\rm A}}$ and
$m_{i,{\rm A}}$, are equal to their expectation values:

\begin{eqnarray}
\label{eq:asimovn}
n_{i,{\rm A}} & = & E[n_i] = \nu_i
= \mu^{\prime} s_i(\vec{\theta}) + b_i(\vec{\theta})  \;, \\*[0.2 cm]
\label{eq:asimovm}
m_{i,{\rm A}} & = & E[m_i] = u_i(\vec{\theta}) \;.
\end{eqnarray}

\noindent Here the parameter values represent those implied by the
assumed distribution of the data.  In practice, these are the values
that would be estimated from the Monte Carlo model using a very large
data sample.

We can use the Asimov data set to evaluate the ``Asimov likelihood''
$L_{\rm A}$ and the corresponding profile likelihood ratio
$\lambda_{\rm A}$.  The use of non-integer values for the data is not
a problem as the factorial terms in the Poisson likelihood represent
constants that cancel when forming the likelihood ratio, and thus can
be dropped.  One finds

\begin{equation}
\label{eq:asimovlambdaii}
\lambda_{\rm A}(\mu)
= \frac{ L_{\rm A}( \mu, \hat{\hat{\vec{\theta}}} ) } { L_{\rm A}(\hat{\mu},
\hat{\vec{\theta}}) } = \frac{ L_{\rm A}( \mu,
\hat{\hat{\vec{\theta}}} ) } { L_{A}(\mu^{\prime}, \vec{\theta} ) } \;,
\end{equation}

\noindent where the final equality above exploits the fact that
the estimators for the parameters are equal to their hypothesized
values when the likelihood is evaluated with the Asimov data set.

A standard way to find $\sigma$ is by estimating the matrix of second
derivatives of the log-likelihood function (cf.\
Eq.~(\ref{eq:vinvwald})) to obtain the inverse covariance matrix
$V^{-1}$, inverting to find $V$, and then extracting the element
$V_{00}$ corresponding to the variance of $\hat{\mu}$.  The second
derivative of $\ln L$ is

\begin{eqnarray}
\label{eq:d2lnl}
\frac{ \partial^2 \ln L }{\partial \theta_j \partial \theta_k } & = &
\sum_{i=1}^N \left[ \left( \frac{n_i}{\nu_i} - 1 \right)
\frac{ \partial^2 \nu_i }{\partial \theta_j \partial \theta_k } -
\frac{ \partial \nu_i }{\partial \theta_j }
\frac{ \partial \nu_i }{\partial \theta_k }
\frac{n_i}{\nu_i^2} \right] \nonumber \\*[0.3 cm]
& + & \sum_{i=1}^M \left[ \left( \frac{m_i}{u_i} - 1 \right)
\frac{ \partial^2 u_i }{\partial \theta_j \partial \theta_k } -
\frac{ \partial u_i }{\partial \theta_j }
\frac{ \partial u_i }{\partial \theta_k }
\frac{m_i}{u_i^2} \right] \;.
\end{eqnarray}

\noindent From (\ref{eq:d2lnl}) one sees that the second derivative of
$\ln L$ is linear in the data values $n_i$ and $m_i$.  Thus its
expectation value is found simply by evaluating with the expectation
values of the data, which is the same as the Asimov data.  One can
therefore obtain the inverse covariance matrix from

\begin{equation}
\label{eq:invcov}
V^{-1}_{jk} = - E \left[ \frac{ \partial^2 \ln L }{\partial \theta_j
\partial \theta_k } \right] = - \frac{ \partial^2 \ln L_{\rm A}
}{\partial \theta_j \partial \theta_k } =
\sum_{i=1}^N \frac{\partial \nu_i}{\partial \theta_j} \frac{\partial
\nu_i}{\partial \theta_k} \frac{1}{\nu_i} + \sum_{i=1}^M
\frac{\partial u_i}{\partial \theta_j} \frac{\partial u_i}{\partial
\theta_k} \frac{1}{u_i} \;.
\end{equation}

\noindent In practice one could, for example, evaluate the the
derivatives of $\ln L_{\rm A}$ numerically, use this to find the
inverse covariance matrix, and then invert and extract the variance of
$\hat{\mu}$.  One can see directly from Eq.~(\ref{eq:invcov}) that
this variance depends on the parameter values assumed for the Asimov
data set, in particular on the assumed strength parameter
$\mu^{\prime}$, which enters via Eq.~(\ref{eq:nui}).

Another method for estimating $\sigma$ (denoted $\sigma_{\rm A}$ in
this section to distinguish it from the approach above based on the
second derivatives of $\ln L$) is to find find the value that is
necessary to recover the known properties of $-\lambda_{\rm A}(\mu)$.
Because the Asimov data set corresponding to a strength $\mu^{\prime}$
gives $\hat{\mu} = \mu^{\prime}$, from Eq.~(\ref{eq:wald}) one finds

\begin{equation}
\label{eq:2lnLambdaAsimov}
- 2 \ln \lambda_{\rm A}(\mu) \approx
\frac{ (\mu - \mu^{\prime} )^2}{\sigma^2} = \Lambda \;.
\end{equation}

\noindent That is, from the Asimov data set one obtains an estimate of
the noncentrality parameter $\Lambda$ that characterizes the
distribution $f(q_{\mu} | \mu^{\prime})$.  Equivalently, one can use
Eq.~(\ref{eq:2lnLambdaAsimov}) to obtain the variance $\sigma^2$ which
characterizes the distribution of $\hat{\mu}$, namely,

\begin{equation}
\label{eq:sigma2}
\sigma_{\rm A}^2 = \frac{(\mu - \mu^{\prime})^2 }{ q_{\mu,{\rm A}} } \;,
\end{equation}

\noindent where $q_{\mu,{\rm A}} = - 2 \ln \lambda_{\rm A}(\mu)$.  For
the important case where one wants to find the median exclusion
significance for the hypothesis $\mu$ assuming that there is no
signal, then one has $\mu^{\prime} = 0$ and therefore

\begin{equation}
\label{eq:sigma2mu}
\sigma_{\rm A}^2 = \frac{\mu^2}{q_{\mu,{\rm A}}} \;,
\end{equation}

\noindent and for the modified statistic $\tilde{q}_{\mu}$ the
analogous relation holds.  For the case of discovery where one tests
$\mu=0$ one has

\begin{equation}
\label{eq:sigma02}
\sigma_{\rm A}^2 = \frac{\mu^{{\prime}\,2}}{q_{0,{\rm A}}} \;.
\end{equation}

The two methods for obtaining $\sigma$ and $\Lambda$ --- from the
Fisher information matrix or from $q_{\mu,\rm A}$ --- are not
identical, but were found to provide similar results in examples of of
practical interest.  In several cases that we considered, the
distribution based on $\sigma_{\rm A}$ provided a better approximation
to the true sampling distribution than the standard approach based on
the Fisher information matrix, leading to the conjecture that it may
effectively incorporate some higher-order terms in
Eq.~(\ref{eq:wald}).

This can be understood qualitatively by noting that under assumption
of the Wald approximation, the test statistics $q_0$, $q_{\mu}$ and
$\tilde{q}_{\mu}$ are monotonically related to $\hat{\mu}$, and
therefore their median values can be found directly by using the
median of $\hat{\mu}$, which is $\mu^{\prime}$.  But monotonicity is a
weaker condition than the full Wald approximation.  That is, even if
higher-order terms are present in Eq.~(\ref{eq:wald}), they will not
alter the distribution's median as long as they do not break the
monotonicity of the relation between the test statistic and
$\hat{\mu}$.  If one uses $\sigma_{\rm A}$ one obtains distributions
with medians given by the corresponding Asimov values, $q_{0,{\rm A}}$
or $q_{\mu,{\rm A}}$, and these values will be correct to the extent
that monotonicity holds.

\subsection{Distribution of $t_{\mu}$}
\label{sec:tmudist}

Consider first using the statistic $t_{\mu} = -2 \ln \lambda(\mu)$ of
Sec.~\ref{sec:tmu} as the basis of the statistical test of a
hypothesized value of $\mu$.  This could be a test of $\mu=0$ for
purposes of establishing existence of a signal process, or non-zero
values of $\mu$ for purposes of obtaining a confidence interval.  To
find the $p$-value $p_{\mu}$, we require the pdf $f(t_{\mu} | \mu)$,
and to find the median $p$-value assuming a different strength
parameter we will need $f(t_{\mu} | \mu^{\prime})$.

The pdf $f(t_{\mu} | \mu^{\prime})$ is given by
Eq.~(\ref{eq:ftmulambda}), namely,

\begin{equation}
\label{eq:ftmumPrime}
f(t_{\mu} | \mu^{\prime}) = \frac{1}{2 \sqrt{t_{\mu}}}
\frac{1}{\sqrt{2 \pi}} \left[ \exp \left( - \frac{1}{2} \left(
\sqrt{t_{\mu}} + \frac{\mu - \mu^{\prime}}{\sigma} \right)^2 \right) +
\exp \left( - \frac{1}{2} \left( \sqrt{t_{\mu}} - \frac{\mu -
\mu^{\prime}}{\sigma} \right)^2 \right) \right] \;.
\end{equation}

\noindent The special case $\mu = \mu^{\prime}$ is simply a chi-square
distribution for one degree of freedom:

\begin{equation}
\label{eq:ftmumu}
f(t_{\mu}|\mu) =
\frac{1}{\sqrt{2 \pi}} \frac{1}{\sqrt{t_{\mu}}}
e^{- t_{\mu} / 2 } \;.
\end{equation}

\noindent
The cumulative distribution of $t_{\mu}$ assuming $\mu^{\prime}$ is

\begin{equation}
\label{eq:tmumuprimecdf}
F(t_{\mu} | \mu^{\prime}) = \Phi \left( \sqrt{t_{\mu}} + \frac{ \mu -
\mu^{\prime} }{\sigma} \right) + \Phi \left( \sqrt{t_{\mu}} - \frac{
\mu - \mu^{\prime} }{\sigma} \right) - 1 \;,
\end{equation}

\noindent where $\Phi$ is the cumulative distribution of the standard
(zero mean, unit variance) Gaussian.  The special case $\mu =
\mu^{\prime}$ is therefore

\begin{equation}
\label{eq:tmumucdf}
F(t_{\mu} | \mu) =
2 \Phi \left( \sqrt{t_{\mu}} \right) - 1 \;,
\end{equation}

\noindent The $p$-value of a hypothesized value of $\mu$
for an observed value $t_{\mu}$ is therefore

\begin{equation}
\label{eq:pmufromtmu}
p_{\mu} = 1 - F(t_{\mu} | \mu) = 2 \left( 1 -
\Phi \left( \sqrt{ t_{\mu} } \right) \right) \;,
\end{equation}

\noindent and the corresponding significance is

\begin{equation}
\label{eq:tmusig}
Z_{\mu} = \Phi^{-1}(1 - p_{\mu}) = \Phi^{-1} \left(
2 \Phi \left( \sqrt{ t_{\mu} } \right) - 1 \right) \;.
\end{equation}

\noindent If the $p$-value is found below a specified threshold
$\alpha$ (often one takes $\alpha = 0.05$), then the value of $\mu$ is
said to be excluded at a confidence level (CL) of $1 - \alpha$.  The
set of points not excluded form a confidence interval with $\mbox{CL}
= 1 - \alpha$.  Here the endpoints of the interval can be obtained
simply by setting $p_{\mu} = \alpha$ and solving for $\mu$.  Assuming
the Wald approximation (\ref{eq:wald}) and using
Eq.~(\ref{eq:pmufromtmu}) one finds

\begin{equation}
\label{eq:mulofromtmu}
\mu_{\rm up/lo} = \hat{\mu} \pm \sigma \Phi^{-1}(1 - \alpha/2) \;.
\end{equation}

\noindent One subtlety with this formula is that $\sigma$ itself
depends at some level on $\mu$.  In practice to find the upper and
lower limits one can simply solve numerically to find those values of
$\mu$ that satisfy $p_{\mu} = \alpha$.

\subsection{Distribution of $\tilde{t}_{\mu}$}
\label{sec:tmutildedist}

Assuming the Wald approximation, the statistic $\tilde{t}_{\mu}$ as
defined by Eq.~(\ref{eq:tmutilde}) can be written

\begin{equation}
\label{eq:tmutildewald}
\tilde{t}_{\mu} =
 \: \left\{ \! \! \begin{array}{ll}
\frac{\mu^2}{\sigma^{2}} - \frac{2 \mu \hat{\mu}}{\sigma^{2}}
                 & \quad \hat{\mu} < 0 \;, \\*[0.2 cm]
\frac{(\mu - \hat{\mu})^2}{\sigma^{2}}
                 &  \quad \hat{\mu} \ge 0  \;.
              \end{array}
       \right.
\end{equation}

\noindent From this the pdf $f(\tilde{t}_{\mu} | \mu^{\prime})$ is
found to be

\begin{eqnarray}
\label{eq:ftildetmmp}
f(\tilde{t}_{\mu}|\mu^{\prime}) & = & \frac{1}{2} \frac{1}{\sqrt{2 \pi}}
\frac{1}{\sqrt{\tilde{t}_{\mu}}} \exp \left[
-\half \left( \sqrt{\tilde{t}_{\mu}} +
\frac{\mu - \mu^{\prime}}{\sigma} \right)^2 \right] \\*[0.3 cm]
& + &
 \: \left\{ \! \!
 \begin{array}{ll}
 \frac{1}{2} \frac{1}{\sqrt{2 \pi}}
\frac{1}{\sqrt{\tilde{t}_{\mu}}} \exp \left[
-\half \left( \sqrt{\tilde{t}_{\mu}} -
\frac{\mu - \mu^{\prime}}{\sigma} \right)^2 \right] &
\quad \tilde{t}_{\mu} \le \mu^2/\sigma^2 \;, \\*[0.5 cm]
\frac{1}{\sqrt{2 \pi}(2\mu/\sigma)} \exp \left[ - \half
\frac{ \left( \tilde{t}_{\mu} -
\frac{\mu^2 - 2 \mu \mu^{\prime}}{\sigma^2} \right)^2 }
{ (2\mu/\sigma)^2 } \right]
 & \quad \tilde{t}_{\mu} > \mu^2/\sigma^2
 \end{array}
 \right.
\;.
\end{eqnarray}

\noindent The special case $\mu = \mu^{\prime}$ is therefore

\begin{equation}
\label{eq:ftildetmm}
f(\tilde{t}_{\mu}|\mu^{\prime}) =
 \: \left\{ \! \!
 \begin{array}{ll}
\frac{1}{\sqrt{2 \pi}}
\frac{1}{\sqrt{\tilde{t}_{\mu}}} e^{- \tilde{t}_{\mu} / 2 }
 & \quad \tilde{t}_{\mu} \le \mu^2/\sigma^2 \;, \\*[0.5 cm]
\frac{1}{2} \frac{1}{\sqrt{2 \pi}}
\frac{1}{\sqrt{\tilde{t}_{\mu}}} e^{- \tilde{t}_{\mu} / 2 } \: + \:
\frac{1}{\sqrt{2 \pi} (2 \mu/\sigma)} \exp \left[ - \half
\frac{ (\tilde{t}_{\mu} + \mu^2/\sigma^2)^2}{ (2 \mu/\sigma)^2 }  \right]
& \quad \tilde{t}_{\mu} > \mu^2/\sigma^2 \;.
 \end{array}
 \right.
\;.
\end{equation}

\noindent The corresponding cumulative distribution is

\begin{equation}
\label{eq:tildetmmpcdf}
F(\tilde{t}_{\mu}|\mu^{\prime}) = \Phi\left( \sqrt{\tilde{t}_{\mu}} +
\frac{\mu - \mu^{\prime}}{\sigma} \right) +
 \: \left\{ \! \! \begin{array}{ll}
\Phi\left( \sqrt{\tilde{t}_{\mu}} -
\frac{\mu - \mu^{\prime}}{\sigma} \right) - 1
                 & \quad \tilde{t}_{\mu} \le \mu^2/\sigma^{2}
\;, \\*[0.5 cm]
\Phi \left( \frac{ \tilde{t}_{\mu} -
(\mu^2 - 2 \mu \mu^{\prime})/\sigma^{2}}
{2\mu/\sigma} \right) - 1
                 &  \quad \tilde{t}_{\mu} > \mu^2/\sigma^{2} \;.
              \end{array}
       \right.
\end{equation}

\noindent For $\mu = \mu^{\prime}$ this is

\begin{equation}
\label{eq:tildetmmcdf}
F(\tilde{t}_{\mu}|\mu) =
 \: \left\{ \! \! \begin{array}{ll}
2 \Phi\Big( \sqrt{\tilde{t}_{\mu}} \Big) - 1
                 & \quad \tilde{t}_{\mu} \le \mu^2/\sigma^2
\;, \\*[0.5 cm]
\Phi\Big( \sqrt{\tilde{t}_{\mu}} \Big) +
\Phi \left( \frac{ \tilde{t}_{\mu} + \mu^2/\sigma^2}
{2\mu/\sigma} \right) - 1
                 &  \quad \tilde{t}_{\mu} > \mu^2/\sigma^2 \;.
              \end{array}
       \right.
\end{equation}

\noindent The $p$-value of the hypothesized $\mu$ is
given by one minus the cumulative distribution, under assumption
of the parameter $\mu$,

\begin{equation}
\label{eq:pvaltmutilde}
p_{\mu} = 1 - F(\tilde{t}_{\mu} | \mu) \;.
\end{equation}

\noindent The corresponding significance is $Z_{\mu} = \Phi^{-1}(1 -
p_{\mu})$.

A confidence interval for $\mu$ at confidence level $\mbox{CL} = 1 -
\alpha$ can be constructed from the set $\mu$ values for which the
$p$-value is not less than $\alpha$.  To find the endpoints of this
interval, one can set $p_{\mu}$ from Eq.~(\ref{eq:pvaltmutilde}) equal
to $\alpha$ and solve for $\mu$.  In general this must be done
numerically.  In the large sample limit, i.e., assuming the validity
of the asymptotic approximations, these intervals correspond to the
limits of Feldman and Cousins \cite{FC} for the case where physical
range of the parameter $\mu$ is $\mu \ge 0$.

\subsection{Distribution of $q_0$ (discovery)}
\label{sec:q0dist}


Assuming the validity of the approximation (\ref{eq:wald}), one has
$-2 \ln \lambda(0) = \hat{\mu}^2 / \sigma^2$.  From the
definition (\ref{eq:q0}) of $q_0$, we therefore have

\begin{equation}
\label{eq:q0wald}
q_{0} =
\left\{ \! \! \begin{array}{ll}
               \hat{\mu}^2 / \sigma^2
               & \quad \hat{\mu} \ge 0 \;, \\*[0.3 cm]
               0 & \quad \hat{\mu} < 0  \;,
              \end{array}
       \right.
\end{equation}

\noindent where $\hat{\mu}$ follows a Gaussian distribution with mean
$\mu^{\prime}$ and standard deviation $\sigma$.  From this one can
show that the pdf of $q_0$ has the form

\begin{equation}
\label{eq:fq0muprimewald}
f(q_0 | \mu^{\prime}) = \left( 1 -
\Phi \left( \frac{ \mu^{\prime}}{\sigma} \right) \right) \delta(q_0)  +
\frac{1}{2}
\frac{1}{\sqrt{2 \pi}} \frac{1}{\sqrt{q_0}} \exp
\left[ - \frac{1}{2} \left( \sqrt{q_0} - \frac{\mu^{\prime}}{\sigma}
\right)^2 \right]
\;.
\end{equation}

\noindent For the special case of $\mu^{\prime} = 0$, this reduces to

\begin{equation}
\label{eq:fq00}
f(q_0 | 0) = \frac{1}{2} \delta(q_0) +
\frac{1}{2} \frac{1}{\sqrt{2 \pi}} \frac{1}{\sqrt{q_0}} e^{-q_0/2} \;.
\end{equation}

\noindent That is, one finds a mixture of a delta function at zero and
a chi-square distribution for one degree of freedom, with each term
having a weight of $1/2$.  In the following we will refer to this
mixture as a half chi-square distribution or $\half \chi^2_1$.

From Eq.~(\ref{eq:fq0muprimewald}) the corresponding cumulative
distribution is found to be

\begin{equation}
\label{cdfq0muprimewald}
F(q_0 | \mu^{\prime}) = \Phi \left( \sqrt{q_0} - \frac{\mu^{\prime}}{\sigma}
\right) \;.
\end{equation}

\noindent The important special case $\mu^{\prime} = 0$ is therefore simply

\begin{equation}
\label{cdfq00wald}
F(q_0 | 0) = \Phi \Big( \sqrt{q_0} \Big)
\;.
\end{equation}

\noindent
The $p$-value of the $\mu=0$ hypothesis (see Eq.~(\ref{eq:q0pval})) is

\begin{equation}
\label{eq:pval0}
p_0 = 1 - F(q_0 | 0) \;,
\end{equation}

\noindent and therefore using Eq.~(\ref{eq:significance})
for the significance one obtains the simple formula

\begin{equation}
\label{eq:Z0}
Z_0 = \Phi^{-1}(1 - p_0) = \sqrt{q_0} \;.
\end{equation}

\subsection{Distribution of $q_{\mu}$ (upper limits)}
\label{sec:qmudist}

Assuming the validity of the Wald approximation, we can write
the test statistic used for upper limits, Eq.~(\ref{eq:qmu}) as

\begin{equation}
\label{eq:qmuwald}
q_{\mu} =
 \quad \: \left\{ \! \! \begin{array}{lll}
               \frac{(\mu - \hat{\mu})^2}{\sigma^{2}}
                & \hat{\mu} < \mu \;, \\*[0.2 cm]
               0 & \hat{\mu} > \mu \;,
              \end{array}
       \right.
\end{equation}

\noindent where $\hat{\mu}$ as before follows a Gaussian centred about
$\mu^{\prime}$ with a standard deviation $\sigma$.

The pdf $f(q_{\mu} | \mu^{\prime})$ is found to be

\begin{equation}
\label{eq:fqmmp}
f(q_{\mu}|\mu^{\prime}) =
\Phi\left( \frac{\mu^{\prime} - \mu}{\sigma} \right)
\delta(q_{\mu}) +
\frac{1}{2} \frac{1}{\sqrt{2 \pi}} \frac{1}{\sqrt{q_{\mu}}}
\exp \left[ -\frac{1}{2} \left( \sqrt{q_{\mu}} -
\frac{\mu - \mu^{\prime}}{\sigma} \right)^2 \right] \;,
\end{equation}

\noindent so that the special case $\mu = \mu^{\prime}$ is
a half-chi-square distribution:

\begin{equation}
\label{eq:fqmm}
f(q_{\mu}|\mu) =
\frac{1}{2} \delta(q_{\mu}) +
\frac{1}{2} \frac{1}{\sqrt{2 \pi}} \frac{1}{\sqrt{q_{\mu}}}
e^{- q_{\mu}/2} \;.
\end{equation}

\noindent The cumulative distribution is

\begin{equation}
\label{eq:qmmpcdf}
F(q_{\mu}|\mu^{\prime}) =
\Phi\left( \sqrt{q_{\mu}} - \frac{\mu - \mu^{\prime}}{\sigma}
\right)
\;,
\end{equation}

\noindent and the corresponding special case $\mu^{\prime} = \mu$ is
thus the same as what was found for $q_0$, namely,

\begin{equation}
\label{eq:qmmcdf}
F(q_{\mu}|\mu) =
\Phi \Big( \sqrt{q_{\mu}} \Big)
\;.
\end{equation}

\noindent The $p$-value of the hypothesized $\mu$ is

\begin{equation}
\label{eq:pvalmu}
p_{\mu} = 1 - F(q_{\mu} |\mu ) = 1 - \Phi \Big( \sqrt{q_{\mu}} \Big)
\end{equation}

\noindent and therefore the corresponding significance is

\begin{equation}
\label{eq:Zmu}
Z_{\mu} = \Phi^{-1}(1 - p_{\mu}) = \sqrt{q_{\mu}} \;.
\end{equation}

\noindent As with the statistic $t_{\mu}$ above,
if the $p$-value is found below a specified
threshold $\alpha$ (often one takes $\alpha = 0.05$), then the value
of $\mu$ is said to be excluded at a confidence level (CL) of $1 -
\alpha$.  The upper limit on $\mu$ is the largest $\mu$ with $p_{\mu}
\le \alpha$.  Here this can be obtained simply by setting $p_{\mu} =
\alpha$ and solving for $\mu$.  Using Eqs.~(\ref{eq:qmuwald}) and
(\ref{eq:pvalmu}) one finds

\begin{equation}
\label{eq:muup}
\mu_{\rm up} = \hat{\mu} + \sigma \Phi^{-1}(1 - \alpha) \;.
\end{equation}

\noindent For example, $\alpha = 0.05$ gives $\Phi^{-1}(1 - \alpha) =
1.64$.  Also as noted above, $\sigma$ depends in general on the
hypothesized $\mu$.  Thus in practice one may find the upper limit
numerically as the value of $\mu$ for which $p_{\mu} = \alpha$.

\subsection{Distribution of $\tilde{q}_{\mu}$ (upper limits)}
\label{sec:qtildemudist}

Using the alternative statistic $\tilde{q}_{\mu}$ defined
by Eq.~(\ref{eq:qmutilde}) and assuming the Wald approximation
we find

\begin{equation}
\label{eq:qtildemuwald}
\tilde{q}_{\mu} =
 \: \left\{ \! \! \begin{array}{lll}
\frac{\mu^2}{\sigma^{2}} - \frac{2 \mu \hat{\mu}}{\sigma^{2}}
                 & \quad \hat{\mu} < 0 \;, \\*[0.2 cm]
\frac{(\mu - \hat{\mu})^2}{\sigma^{2}}
                 &  \quad 0 \le \hat{\mu} \le \mu  \;, \\*[0.2 cm]
               0 & \quad \hat{\mu} > \mu \;.
              \end{array}
       \right.
\end{equation}

\noindent The pdf $f(\tilde{q}_{\mu} | \mu^{\prime})$
is found to be

\begin{eqnarray}
\label{eq:ftildeqmmp}
f(\tilde{q}_{\mu}|\mu^{\prime}) & = &
\Phi \left( \frac{\mu^{\prime} - \mu}{\sigma} \right)
\delta (\tilde{q}_{\mu}) \nonumber \\*[0.3 cm]
& + &
 \: \left\{ \! \! \begin{array}{lll}
\frac{1}{2} \frac{1}{\sqrt{2 \pi}} \frac{1}{\sqrt{\tilde{q}_{\mu}}}
\exp \left[ -\frac{1}{2} \left( \sqrt{\tilde{q}_{\mu}} -
\frac{\mu - \mu^{\prime}}{\sigma} \right)^2 \right]
                 & 0 < \tilde{q}_{\mu} \le \mu^2/\sigma^{2}  \;, \\*[0.5 cm]
\frac{1}{\sqrt{2 \pi} (2\mu/\sigma)} \exp \left[
-\frac{1}{2} \frac{ (\tilde{q}_{\mu} -
(\mu^2 - 2 \mu \mu^{\prime})/\sigma^{2} )^2 }
{(2 \mu/\sigma)^2} \right]
                 &  \quad \tilde{q}_{\mu} > \mu^2/\sigma^{2}  \;.
              \end{array}
       \right.
\end{eqnarray}

\noindent
The special case $\mu = \mu^{\prime}$ is therefore

\begin{equation}
\label{eq:ftildeqmm}
f(\tilde{q}_{\mu}|\mu) =
\frac{1}{2} \delta (\tilde{q}_{\mu}) +
 \: \left\{ \! \! \begin{array}{lll}
\frac{1}{2} \frac{1}{\sqrt{2 \pi}} \frac{1}{\sqrt{\tilde{q}_{\mu}}}
e^{- \tilde{q}_{\mu}/2}
                 & 0 < \tilde{q}_{\mu} \le \mu^2/\sigma^2  \;, \\*[0.5 cm]
\frac{1}{\sqrt{2 \pi} (2\mu/\sigma)} \exp \left[
-\frac{1}{2} \frac{ (\tilde{q}_{\mu} + \mu^2/\sigma^2 )^2 }
{(2 \mu/\sigma)^2} \right]
                 &  \quad \tilde{q}_{\mu} > \mu^2/\sigma^2 \;.
              \end{array}
       \right.
\end{equation}

\noindent The corresponding cumulative distribution is

\begin{equation}
\label{eq:tildeqmmpcdf}
F(\tilde{q}_{\mu}|\mu^{\prime}) =
 \: \left\{ \! \! \begin{array}{lll}
\Phi\left( \sqrt{\tilde{q}_{\mu}} -
\frac{\mu - \mu^{\prime}}{\sigma} \right)
                 & \quad 0 < \tilde{q}_{\mu} \le \mu^2/\sigma^{2}
\;, \\*[0.5 cm]
\Phi \left( \frac{ \tilde{q}_{\mu} -
(\mu^2 - 2 \mu \mu^{\prime})/\sigma^{2}}
{2\mu/\sigma} \right)
                 &  \quad \tilde{q}_{\mu} > \mu^2/\sigma^{2} \;.
              \end{array}
       \right.
\end{equation}

\noindent The special case $\mu = \mu^{\prime}$ is

\begin{equation}
\label{eq:tildeqmmcdf}
F(\tilde{q}_{\mu}|\mu) =
 \: \left\{ \! \! \begin{array}{lll}
\Phi\Big( \sqrt{\tilde{q}_{\mu}} \Big)
                 & \quad 0 < \tilde{q}_{\mu} \le \mu^2/\sigma^2
\;, \\*[0.5 cm]
\Phi \left( \frac{ \tilde{q}_{\mu} + \mu^2/\sigma^2}
{2\mu/\sigma} \right)
                 &  \quad \tilde{q}_{\mu} > \mu^2/\sigma^2 \;.
              \end{array}
       \right.
\end{equation}

\noindent The $p$-value of the hypothesized $\mu$ is as before
given by one minus the cumulative distribution,

\begin{equation}
\label{eq:pvalmutilde}
p_{\mu} = 1 - F(\tilde{q}_{\mu} | \mu) \;,
\end{equation}

\noindent and therefore the corresponding significance is

\begin{equation}
\label{eq:zmutilde}
Z_{\mu} =
 \: \left\{ \! \! \begin{array}{lll}
\sqrt{\tilde{q}_{\mu}}
     & \quad 0 < \tilde{q}_{\mu} \le \mu^2/\sigma^2  \;, \\*[0.5 cm]
\frac{ \tilde{q}_{\mu} + \mu^2/\sigma^2}{2\mu/\sigma}
                 &  \quad \tilde{q}_{\mu} > \mu^2/\sigma^2 \;.
              \end{array}
       \right.
\end{equation}

As when using $q_{\mu}$, the upper limit on $\mu$ at confidence level
$1 - \alpha$ is found by setting $p_{\mu} = \alpha$ and solving for
$\mu$, which reduces to the same result as found when using $q_{\mu}$,
namely,

\begin{equation}
\label{eq:muuptilde}
\mu_{\rm up} =  \hat{\mu} + \sigma \Phi^{-1}(1 - \alpha) \;.
\end{equation}

\noindent That is, to the extent that the Wald approximation holds,
the two statistics $q_{\mu}$ and $\tilde{q}_{\mu}$ lead to identical
upper limits.

\subsection{Distribution of $-2 \ln (L_{s+b}/L_b)$}
\label{sec:tevatron}

Many analyses carried out at the Tevatron Collider (e.g.,
\cite{TevatronSearch}) involving searches for a new signal process
have been based on the statistic

\begin{equation}
\label{eq:qtevdef}
q = -2 \ln \frac{L_{s+b}}{L_b} \;,
\end{equation}

\noindent where $L_{s+b}$ is the likelihood of the nominal signal
model and $L_b$ is that of the background-only hypothesis.  That is,
the $s+b$ corresponds to having the strength parameter $\mu= 1$ and
$L_b$ refers to $\mu = 0$.  The statistic $q$ can therefore be written

\begin{equation}
\label{eq:qtev2}
q = -2 \ln \frac{L(\mu=1, \hat{\hat{\vec{\theta}}}(1))}
{L(\mu=0, \hat{\hat{\vec{\theta}}}(0))}
= - 2 \ln \lambda(1) + 2 \ln \lambda(0) \;.
\end{equation}

Assuming the validity of the Wald approximation (\ref{eq:wald}),
$q$ is given by

\begin{equation}
\label{eq:qtevwald}
q = \frac{(\hat{\mu} - 1)^2 }{\sigma^2}  - \frac{\hat{\mu}^2}{ \sigma^2 }
= \frac{ 1 - 2 \hat{\mu}}{\sigma^2} \;,
\end{equation}

\noindent where as previously $\sigma^2$ is the variance of
$\hat{\mu}$.  As $\hat{\mu}$ follows a Gaussian distribution, the
distribution of $q$ is also seen to be Gaussian, with a mean value of

\begin{equation}
\label{eq:meanqtev}
E[q] = \frac{ 1 - 2 \mu }{\sigma^2}
\end{equation}

\noindent and a variance of

\begin{equation}
\label{eq:varqtev}
V[q] = \frac{4}{\sigma^2} \;.
\end{equation}

\noindent That is, the standard deviation of $q$ is $\sigma_q = 2 /
\sigma$, where the standard deviation of $\hat{\mu}$, $\sigma$, can be
estimated, e.g., using the second derivatives of the log-likelihood
function as described in Sec.~\ref{sec:wald} or with the methods
discussed in Sec.~\ref{sec:asimov}.  Recall that in general $\sigma$
depends on the hypothesized value of $\mu$; here we will refer to
these as $\sigma_{b}$ and $\sigma_{s+b}$ for the $\mu=0$ and $\mu=1$
hypotheses, respectively.

From Eq.~(\ref{eq:meanqtev}) one sees that for the $s+b$ hypothesis
($\mu=1$) the values of $q$ tend to be lower, and for the $b$
hypothesis ($\mu=0$) they are higher.  Therefore we can find the
$p$-values for the two hypothesis from

\begin{eqnarray}
\label{eq:psb} p_{s+b} & = & \int_{q_{\rm obs}}^{\infty} f(q|s+b) \,
dq = 1 - \Phi \left( \frac{q_{\rm obs} +
1/\sigma_{s+b}^2}{2/\sigma_{s+b}} \right) \;,
\\*[0.2 cm]
\label{eq:pb} p_{b} & = & \int_{-\infty}^{q_{\rm obs}} f(q|b) \, dq
= \Phi \left( \frac{q_{\rm obs} - 1/\sigma_b^2}{2/\sigma_b} \right)
\;,
\end{eqnarray}

\noindent where we have used Eqs.~(\ref{eq:meanqtev}) and
(\ref{eq:varqtev}) for the mean and variance of $q$ under the $b$ and
$s+b$ hypotheses.

The $p$-values from Eqs.~(\ref{eq:psb}) and (\ref{eq:pb}) incorporate
the effects of systematic uncertainties to the extent that these are
connected to the nuisance parameters $\vec{\theta}$.  In analyses done
at the Tevatron such as in Ref.~\cite{TevatronSearch}, these effects
are incorporated into the distribution of $q$ in a different but
largely equivalent way.  There, usually one treats the control
measurements that constrain the nuisance parameters as fixed, and to
determine the distribution of $q$ one only generates the main search
measurement (i.e., what corresponds in our generic analysis to the
histogram $\vec{n}$).  The effects of the systematic uncertainties are
taken into account by using the control measurements as the basis of a
Bayesian prior density $\pi(\vec{\theta})$, and the distribution of
$q$ is computed under assumption of the Bayesian model average

\begin{equation}
\label{eq:bayesmodave}
f(q) = \int f(q | \vec{\theta}) \pi(\vec{\theta}) \, d \vec{\theta}
\;.
\end{equation}


The prior pdf $\pi(\vec{\theta})$ used in Eq.~(\ref{eq:bayesmodave})
would be obtained from some measurements characterized by a likelihood
function $L_{\vec{\theta}}(\vec{\theta})$, and then used to find the
prior $\pi(\vec{\theta})$ using Bayes' theorem,

\begin{equation}
\label{eq:bayesthm}
\pi(\vec{\theta}) \propto L_{\vec{\theta}}(\vec{\theta})
\pi_0(\vec{\theta}) \;.
\end{equation}

\noindent Here $\pi_0(\vec{\theta})$ is the initial prior for
$\vec{\theta}$ that reflected one's knowledge before carrying out the
control measurements.  In many cases this is simply take as a
constant, in which case $\pi(\vec{\theta})$ is simply proportional to
$L_{\vec{\theta}}(\vec{\theta})$.

In the approach of this paper, however, all measurements are regarded
as part of the data, including control measurements that constrain
nuisance parameters.  That is, here to generate a data set by MC
means, for a given assumed point in the model's parameter space, one
simulates both the control measurements and the main measurement.
Although this is done for a specific value of $\vec{\theta}$, in the
asymptotic limit the distributions required for computing the
$p$-values (\ref{eq:psb}) and (\ref{eq:pb}) are only weakly dependent
on $\vec{\theta}$ to the extent that this can affect the standard
deviation $\sigma_q$.  By contrast, in the Tevatron approach one
generates only the main measurement with data distributed according to
the averaged model (\ref{eq:bayesmodave}).  In the case where the
nuisance parameters are constrained by Gaussian distributed estimates
and the initial prior $\pi_0(\vec{\theta})$ is taken to be constant,
the two methods are essentially equivalent.

Assuming the Wald approximation holds, the statistic $q$ as well as
$q_0$ from Eq.~(\ref{eq:q0}), $q_{\mu}$ from Eq.~(\ref{eq:qmu}) and
$\tilde{q}_{\mu}$ from Eq.~(\ref{eq:qmutilde}) are all monotonic
functions of $\hat{\mu}$, and therefore all are equivalent to
$\hat{\mu}$ in terms of yielding the same statistical test.  If there
are no nuisance parameters, then the Neyman--Pearson lemma (see,
e.g., \cite{Kendall2}) states that the likelihood ratio
$L_{s+b}/L_{b}$ (or equivalently $q$) is an optimal test statistic in
the sense that it gives the maximum power for a test of the
background-only hypothesis with respect to the alternative of signal
plus background (and vice versa).  But if the Wald approximation
holds, then $q_0$ and $q_{\mu}$ lead to equivalent tests and are
therefore also optimal in the Neyman--Pearson sense.  If the nuisance
parameters are well constrained by control measurements, then one
expects this equivalence to remain approximately true.

Finally, note that in many analyses carried out at the Tevatron,
hypothesized signal models are excluded based not on whether the
$p$-value $p_{s+b}$ from Eq.~(\ref{eq:psb}) is less than a given
threshold $\alpha$, but rather the ratio
%
%
$\mbox{CL}_s = p_{s+b}/(1 - p_{b})$
is compared to $\alpha$.  We do not consider this final step
here; it is discussed in, e.g., Ref.~\cite{cls}.

\section{Experimental sensitivity}
\label{sec:sensitivity}

To characterize the sensitivity of an experiment, one is interested
not in the significance obtained from a single data set, but rather in
the expected (more precisely, median) significance with which one
would be able to reject different values of $\mu$.  Specifically, for
the case of discovery one would like to know the median, under the
assumption of the nominal signal model ($\mu=1$), with which one would
reject the background-only ($\mu=0$) hypothesis.  And for the case of
setting exclusion limits the sensitivity is characterized by the
median significance, assuming data generated using the $\mu=0$
hypothesis, with which one rejects a nonzero value of $\mu$ (usually
$\mu=1$ is of greatest interest).

The sensitivity of an experiment is illustrated in
Fig.~\ref{fig:sensitivity}, which shows the pdf for $q_{\mu}$ assuming
both a strength parameter $\mu$ and also assuming a different value
$\mu^{\prime}$.  The distribution $f(q_{\mu}|\mu^{\prime})$ is
shifted to higher value of $q_{\mu}$, corresponding on average to
lower $p$-values.  The sensitivity of an experiment can be
characterized by giving the $p$-value corresponding to the median
$q_{\mu}$ assuming the alternative value $\mu^{\prime}$.  As the
$p$-value is a monotonic function of $q_{\mu}$, this is equal to the
median $p$-value assuming $\mu^{\prime}$.

\setlength{\unitlength}{1.0 cm}
\renewcommand{\baselinestretch}{0.8}
\begin{figure}[htbp]
\begin{picture}(10.0,5)
\put(1,-0.2){\includegraphics{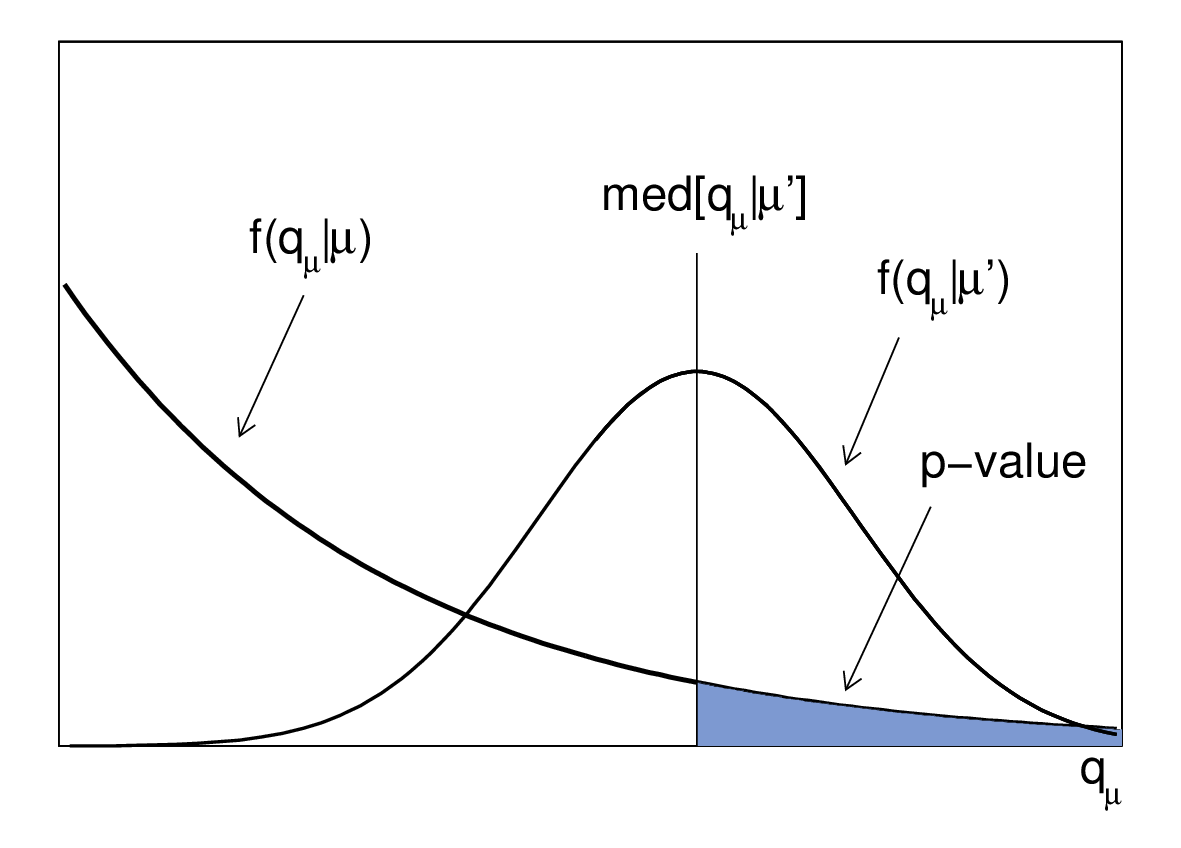}}
\put(9.0,0.){\makebox(5,4)[b]{\begin{minipage}[b]{5cm}
\protect\caption{{\footnotesize Illustration of the
the $p$-value corresponding to the median
of $q_{\mu}$ assuming a strength parameter $\mu^{\prime}$ (see text).}
\protect\label{fig:sensitivity}}
\end{minipage}}}
\end{picture}
\end{figure}
\renewcommand{\baselinestretch}{1}
\small\normalsize

In the rest of this section we describe the ingredients needed to
determine the experimental sensitivity (median discovery or exclusion
significance).  In Sec.~\ref{sec:asimov} we introduced the Asimov data
set, in which all statistical fluctuations are suppressed.  This will
lead directly to estimates of the experimental sensitivity
(Sec.~\ref{sec:medsig}) as well as providing an alternative estimate
of the standard deviation $\sigma$ of the estimator $\hat{\mu}$.  In
Sec.~\ref{sec:multichan} we indicate how the procedure can be extended
to the case where several search channels are combined, and in
Sec.~\ref{sec:statvar} we describe how to give statistical error bands
for the sensitivity.

\subsection{The median significance from Asimov values of the test statistic}
\label{sec:medsig}


By using the Asimov data set one can easily obtain the median values
of $q_0$, $q_{\mu}$ and $\tilde{q}_{\mu}$, and these lead to simple
expressions for the corresponding median significance.  From
Eqs.~(\ref{eq:Z0}), (\ref{eq:Zmu}) and (\ref{eq:zmutilde}) one sees
that the significance $Z$ is a monotonic function of $q$, and
therefore the median $Z$ is simply given by the corresponding function
of the median of $q$, which is approximated by its Asimov value.  For
discovery using $q_0$ one wants the median discovery significance
assuming a strength parameter $\mu^{\prime}$ and for upper limits one
is particularly interested in the median exclusion significance
assuming $\mu^{\prime} = 0$, $\mbox{med}[Z_{\mu} | 0]$.  For these one
obtains

\begin{eqnarray}
\label{eq:medZ0}
\mbox{med}[Z_0| \mu^{\prime}] & = &
\sqrt{q_{0,{\rm A}}} \;, \\*[0.2 cm]
\label{eq:medZmu}
\mbox{med}[Z_{\mu} | 0 ] & = & \sqrt{q_{\mu,{\rm A}}} \;.
\end{eqnarray}

When using $\tilde{q}_{\mu}$ for establishing upper limits, the
general expression for the exclusion significance $Z_{\mu}$ is
somewhat more complicated depending on $\mu^{\prime}$, but is in any
case found by substituting the appropriate values of
$\tilde{q}_{\mu,{\rm A}}$ and $\sigma_{\rm A}$ into
Eq.~(\ref{eq:zmutilde}).  For the usual case where one wants the
median significance for $\mu$ assuming data distributed according to
the background-only hypothesis ($\mu^{\prime} = 0$),
Eq.~(\ref{eq:zmutilde}) reduces in fact to a relation of the same form
as Eq.~(\ref{eq:Zmu}), and therefore one finds

\begin{equation}
\label{eq:zmutilde2}
\mbox{med}[Z_{\mu}|0] = \sqrt{\tilde{q}_{\mu,{\rm A}}} \;.
\end{equation}

\subsection{Combining multiple channels}
\label{sec:multichan}

In many analyses, there can be several search channels which need to
be combined.  For each channel $i$ there is a likelihood function
$L_i(\mu, \vec{\theta}_i)$, where $\vec{\theta}_i$ represents the set
of nuisance parameters for the $i$th channel, some of which may be
common between channels.  Here the strength parameter $\mu$
is assumed to be the same for all channels.  If the channels are
statistically independent, as can usually be arranged, the full
likelihood function is given by the product over all of the channels,

\begin{equation}
\label{eq:Lfull} L(\mu, \vec{\theta}) = \prod_i L_i (\mu,
\vec{\theta}_i) \;,
\end{equation}

\noindent where $\vec{\theta}$ represents the complete set of all
nuisance parameters.  The profile likelihood ratio $\lambda(\mu)$
is therefore

\begin{equation}
\label{eq:lambdaFull} \lambda(\mu) =
\frac{ \prod_i L_i ( \mu, \hat{\hat{\vec{\theta}}}_i) }
{ \prod_i L_i (\hat{\mu}, \hat{\vec{\theta}}_{i} ) }
\;.
\end{equation}

Because the Asimov data contain no statistical fluctuations, one has
$\hat{\mu} = \mu^{\prime}$ for all channels.  Furthermore any common
components of $\vec{\theta}_{i}$ are the same for all channels.
Therefore when using the Asimov data corresponding to a strength
parameter $\mu^{\prime}$ one finds

\begin{equation}
\label{eq:lambdaAsimovCombo}
\lambda_{\rm A}(\mu) =
\frac{ \prod_i L_i ( \mu, \hat{\hat{\vec{\theta}}}) }
{ \prod_i L_i (\mu^{\prime}, \vec{\theta} ) }
 = \prod_i \lambda_{{\rm A},i}(\mu) \;,
\end{equation}

%

\noindent where $\lambda_{{\rm A},i}(\mu)$ is the profile likelihood
ratio for the $i$th channel alone.


Because of this, it is possible to determine the values of the profile
likelihood ratio entering into (\ref{eq:lambdaAsimovCombo}) separately
for each channel, which simplifies greatly the task of estimating the
median significance that would result from the full combination.  It
should be emphasized, however, that to find the discovery significance
or exclusion limits determined from real data, one needs to construct
the full likelihood function containing a single parameter $\mu$, and
this must be used in a global fit to find the profile likelihood
ratio.


\subsection{Expected statistical variation (error bands)}
\label{sec:statvar}

By using the Asimov data set we can find the median, assuming some
strength parameter $\mu^{\prime}$ of the significance for rejecting a
hypothesized value $\mu$.  Even if the hypothesized value
$\mu^{\prime}$ is correct, the actual data will contain statistical
fluctuations and thus the observed significance is not in general
equal to the median.

For example, if the signal is in fact absent but the number of
background events fluctuates upward, then the observed upper limit on
the parameter $\mu$ will be weaker than the median assuming
background only.  It is useful to know by how much the significance is
expected to vary, given the expected fluctuations in the data.
As we have formulae for all of the relevant sampling distributions,
we can also predict how the significance is expected to vary under
assumption of a given signal strength.

It is convenient to calculate error bands for the median significance
corresponding to the $\pm N \sigma$ variation of $\hat{\mu}$.  As
$\hat{\mu}$ is Gaussian distributed, these error bands on the
significance are simply the quantiles that map onto the variation of
$\hat{\mu}$ of $\pm N \sigma$ about $\mu^{\prime}$.

For the case of discovery, i.e., a test of $\mu = 0$, one has from
Eqs.~(\ref{eq:q0wald}) and (\ref{eq:Z0}) that the significance $Z_0$
is

\begin{equation}
\label{eq:Z0wald}
Z_0 =
\left\{ \! \! \begin{array}{ll}
               \hat{\mu} / \sigma
               & \quad \hat{\mu} \ge 0 \;, \\*[0.3 cm]
               0 & \quad \hat{\mu} < 0  \;.
              \end{array}
       \right.
\end{equation}

\noindent Furthermore the median significance is found from
Eq.~(\ref{eq:medZ0}), so the significance values corresponding to
$\mu^{\prime} \pm N \sigma$ are therefore

\begin{eqnarray}
\label{eq:Z0band}
Z_0(\mu^{\prime}+N\sigma) & = &
\mbox{med}[Z | \mu^{\prime}] + N \;, \\*[0.2 cm]
Z_0(\mu^{\prime}-N\sigma) & = & \mbox{max}\left[
\mbox{med}[Z | \mu^{\prime}] - N, 0 \right] \;.
\end{eqnarray}

For the case of exclusion, when using both the statistic $q_{\mu}$ as
well as $\tilde{q}_{\mu}$ one found the same expression for the upper
limit at a confidence level of $1 - \alpha$, namely,
Eq.~(\ref{eq:muup}).  Therefore the median upper limit assuming a
strength parameter $\mu^{\prime}$ is found simply by substituting this
for $\hat{\mu}$, and the $\pm N \sigma$ error bands are found
similarly by substituting the corresponding values of $\mu^{\prime}
\pm N \sigma$.  That is, the median upper limit is

\begin{equation}
\label{eq:medmuup}
\mbox{med}[\mu_{\rm up}|\mu^{\prime}] =
\mu^{\prime} + \sigma \Phi^{-1}(1 - \alpha) \;,
\end{equation}

\noindent and the $\pm N\sigma$ error band is given by

\begin{equation}
\label{eq:muupband}
\mbox{band}_{N\sigma} =
\mu^{\prime} + \sigma ( \Phi^{-1}(1 - \alpha) \pm N ) \;.
\end{equation}

\noindent  The standard deviation $\sigma$ of $\hat{\mu}$ can
be obtained from the Asimov value of the test statistic $q_{\mu}$
(or $\tilde{q}_{\mu}$) using Eq.~(\ref{eq:sigma2}).

\section{Examples}
\label{sec:examples}

In this section we describe two examples, both of which are special
cases of the generic analysis described in
Section~\ref{sec:formalism}.  Here one has a histogram $\vec{n} =
(n_1, \ldots, n_N)$ for the main measurement where signal events could
be present and one may have another histogram $\vec{m} = (m_1, \ldots,
m_M)$ as a control measurement, which helps constrain the nuisance
parameters.  In Section~\ref{sec:counting} we treat the simple case
where each of these two measurements consists of a single Poisson
distributed value, i.e., the histograms each have a single bin.  We
refer to this as a ``counting experiment''.  In
Section~\ref{sec:shape} we consider multiple bins for the main
histogram, but without a control histogram; here the measured shape of
the main histogram on either side of the signal peak is sufficient to
constrain the background.  We refer to this as a ``shape analysis''.

\subsection{Counting experiment}
\label{sec:counting}

Consider an experiment where one observes a number of events $n$,
assumed to follow a Poisson distribution with an expectation value
$E[n] = \mu s + b$.  Here $s$ represents the mean number of events
from a signal model, which we take to be a known value;
$b$ is the expected number from background processes, and as usual
$\mu$ is the strength parameter.

We will treat $b$ as a nuisance parameter whose value is constrained
by a control measurement.  This measurement is also a single Poisson
distributed value $m$ with mean value $E[m] = \tau b$.  That is, $\tau
b$ plays the role of the function $u$ for the single bin of the
control histogram in Eq.~(\ref{eq:emi}).  In a real analysis, the
value of the scale factor $\tau$ may have some uncertainty and could
be itself treated as a nuisance parameter, but in this example we will
take its value to be known.  Related aspects of this type of analysis
have been discussed in the literature, where it is sometimes referred
to as the ``on-off problem'' (see, e.g., \cite{Cranmer03,Cousins08}).

The data thus consist of two measured values: $n$ and $m$.  We have
one parameter of interest, $\mu$, and one nuisance parameter, $b$.
The likelihood function for $\mu$ and $b$ is the product of two
Poisson terms:

\begin{equation}
\label{eq:likelihoodcounting}
L(\mu, b) = \frac{(\mu s + b)^n}{n!} e^{-(\mu s + b)} \,
\frac{(\tau b)^m}{m!} e^{-\tau b} \;.
\end{equation}

%
%

To find the test statistics $q_0$, $q_{\mu}$ and $\tilde{q}_{\mu}$, we
require the ML estimators $\hat{\mu}$, $\hat{b}$ as well as the
conditional ML estimator $\hat{\hat{b}}$ for a specified $\mu$.  These
are found to be

\begin{eqnarray}
\label{eq:muhatcounting}
\hat{\mu} & = & \frac{n - m/\tau}{s} \;, \\*[0.2 cm]
\label{eq:bhatcounting}
\hat{b} & = & \frac{m}{\tau} \;, \\*[0.2 cm]
\label{eq:bhathatcounting}
\hat{\hat{b}} & = & \frac{n + m - (1 + \tau) \mu s}{2 (1 + \tau)}
+ \left[ \frac{ (n + m - (1+\tau) \mu s)^2 + 4 (1 + \tau) m \mu s}{
4 (1 + \tau)^2 } \right]^{1/2} \;.
\end{eqnarray}

Given measured values $n$ and $m$, the estimators from
Eqs.~(\ref{eq:muhatcounting}), (\ref{eq:bhatcounting}) and
(\ref{eq:bhathatcounting}) can be used in the likelihood function
(\ref{eq:likelihoodcounting}) to find the values of the test
statistics $q_0$, $q_{\mu}$ and $\tilde{q}_{\mu}$.  By generating data
values $n$ and $m$ by Monte Carlo we can compare the resulting
distributions with the formulae from Section~\ref{sec:qdist}.

The pdf $f(q_0|0)$, i.e., the distribution of $q_0$ for under the
assumption of $\mu=0$, is shown in Fig.~\ref{fig:q0counting}(a).  The
histograms show the result from Monte Carlo simulation based on
several different values of the mean background $b$.  The solid curve
shows the prediction of Eq.~(\ref{eq:fq00}), which is independent of
the nuisance parameter $b$.
The point at which one finds a significant departure between the
histogram and the asymptotic formula occurs at increasingly large
$q_0$ for increasing $b$.  For $b=20$ the agreement is already quite
accurate past $q_0 = 25$, corresponding to a significance of $Z =
\sqrt{q_0} = 5$.  Even for $b = 2$ there is good agreement out to $q_0
\approx 10$.

\setlength{\unitlength}{1.0 cm}
\renewcommand{\baselinestretch}{0.9}
\begin{figure}[htbp]
\begin{picture}(10.0,6.5)
\put(.8,0)
{\includegraphics{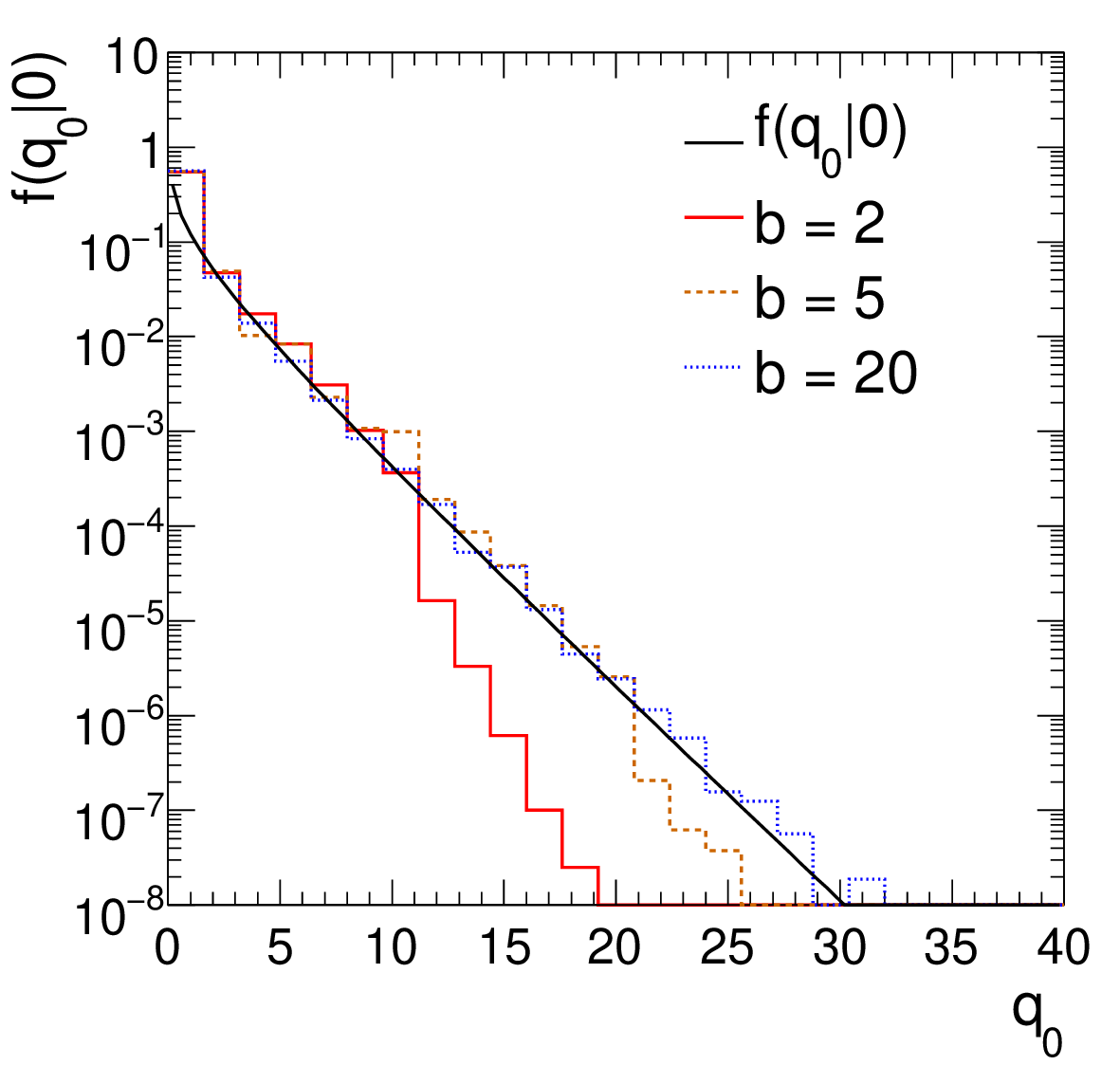}}
\put(8,0)
{\includegraphics{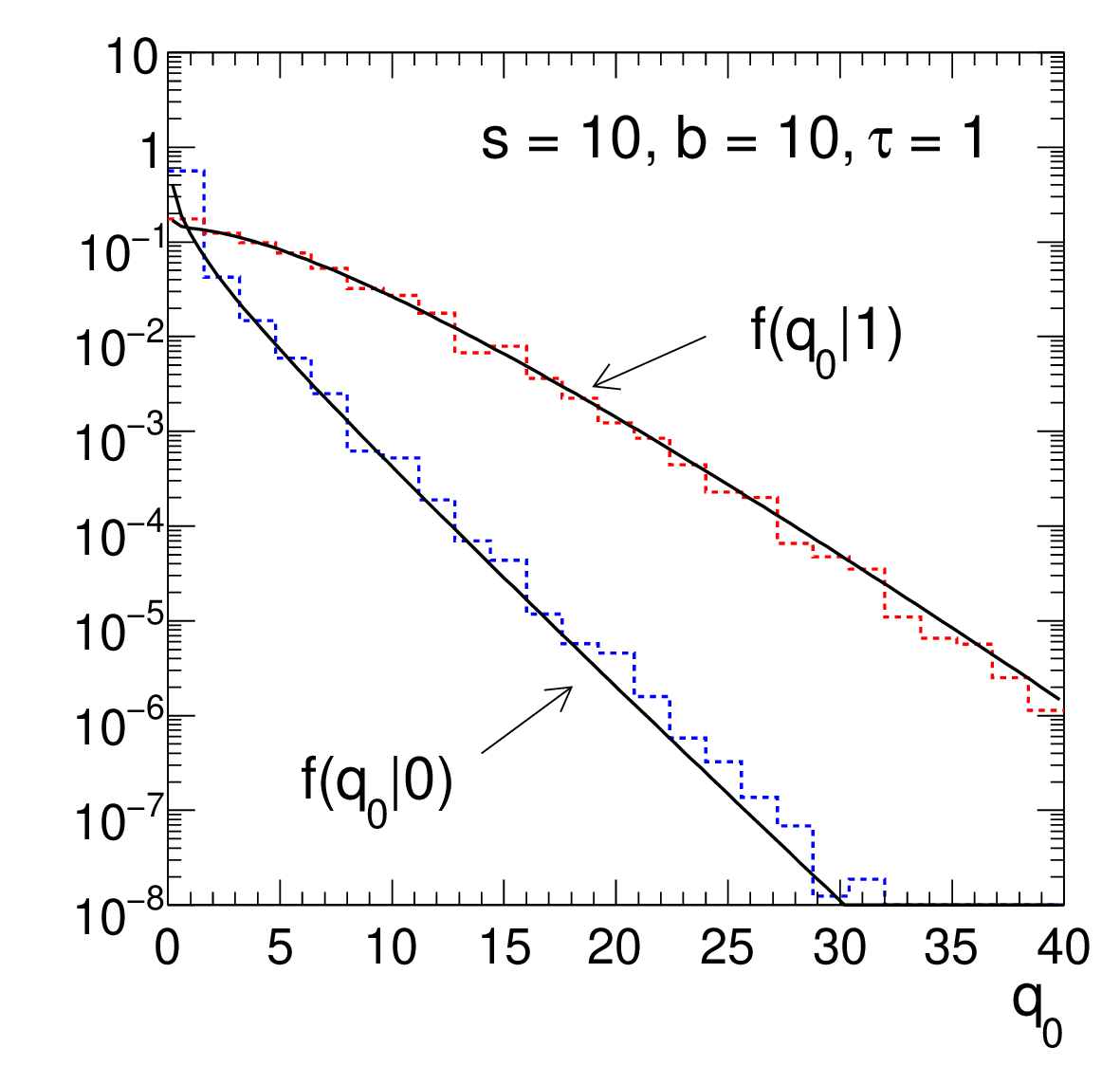}}
\put(0.,6.){(a)}
\put(15,6.){(b)}
\end{picture}
\caption{\small (a) The pdf $f(q_0|0)$ for the
counting experiment.  The solid curve shows $f(q_0|0)$ from
Eq.~(\ref{eq:fq00}) and the histograms are from Monte Carlo using
different values of $b$ (see text).  (b)  The distributions
$f(q_0|0)$ and $f(q_0|1)$ from both the asymptotic formulae and
Monte Carlo simulation based on $s = 10$, $b = 10$, $\tau = 1$.}
\label{fig:q0counting}
\end{figure}
\renewcommand{\baselinestretch}{1}
\small\normalsize

Figure~\ref{fig:q0counting}(b) shows distributions of $q_0$ assuming a
strength parameter $\mu^{\prime}$ equal to 0 and 1.  The histograms
show the Monte Carlo simulation of the corresponding distributions
using the parameters $s = 10$, $b = 10$, $\tau = 1$.  For the
distribution $f(q_0|1)$ from Eq.~(\ref{eq:fq0muprimewald}), one
requires the value of $\sigma$, the standard deviation of $\hat{\mu}$
assuming a strength parameter $\mu^{\prime} = 1$.  Here this was
determined from Eq.~(\ref{eq:sigma02}) using the Asimov value
$q_{0,{\rm A}}$, i.e., the value obtained from the Asimov data set
with $n \rightarrow \mu^{\prime} s + b$ and $m \rightarrow \tau b$.

We can investigate the accuracy of the approximations used by
comparing the discovery significance for a given observed value of
$q_0$ from the approximate formula with the exact significance
determined using a Monte Carlo calculation.
Figure~\ref{fig:q0test}(a) shows the discovery significance that one
finds from $q_0 = 16$.  According to Eq.~(\ref{eq:Z0}), this should
give a nominal significance of $Z = \sqrt{q_0} = 4$, indicated in the
figure by the horizontal line.  The points show the exact significance
for different values of the expected number of background events $b$
in the counting analysis with a scale factor $\tau = 1$.  As can be
seen, the approximation underestimates the significance for very low
$b$, but achieves an accuracy of better than 10\% for $b$ greater than
around 4.  It slightly overestimates for $b$ greater than around 5.
This phenomenon can be seen in the tail of $f(q_0|0)$ in
Fig.~\ref{fig:q0counting}(b), which uses $b = 10$.  The accuracy then
rapidly improves for increasing $b$.

\setlength{\unitlength}{1.0 cm}
\renewcommand{\baselinestretch}{0.9}
\begin{figure}[htbp]
\begin{picture}(10.0,6.5)
\put(.8,0)
{\includegraphics{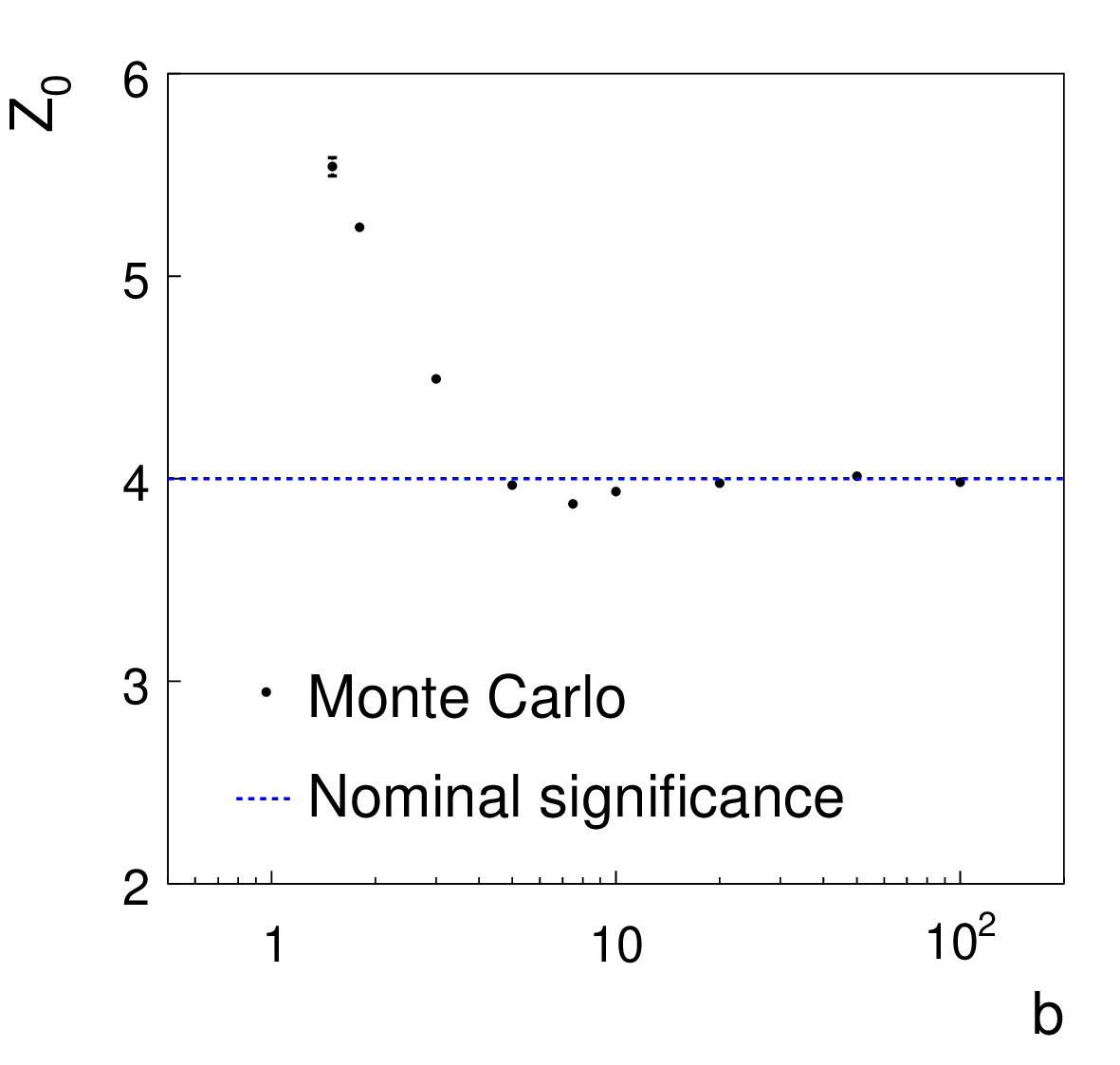}}
\put(8,0)
{\includegraphics{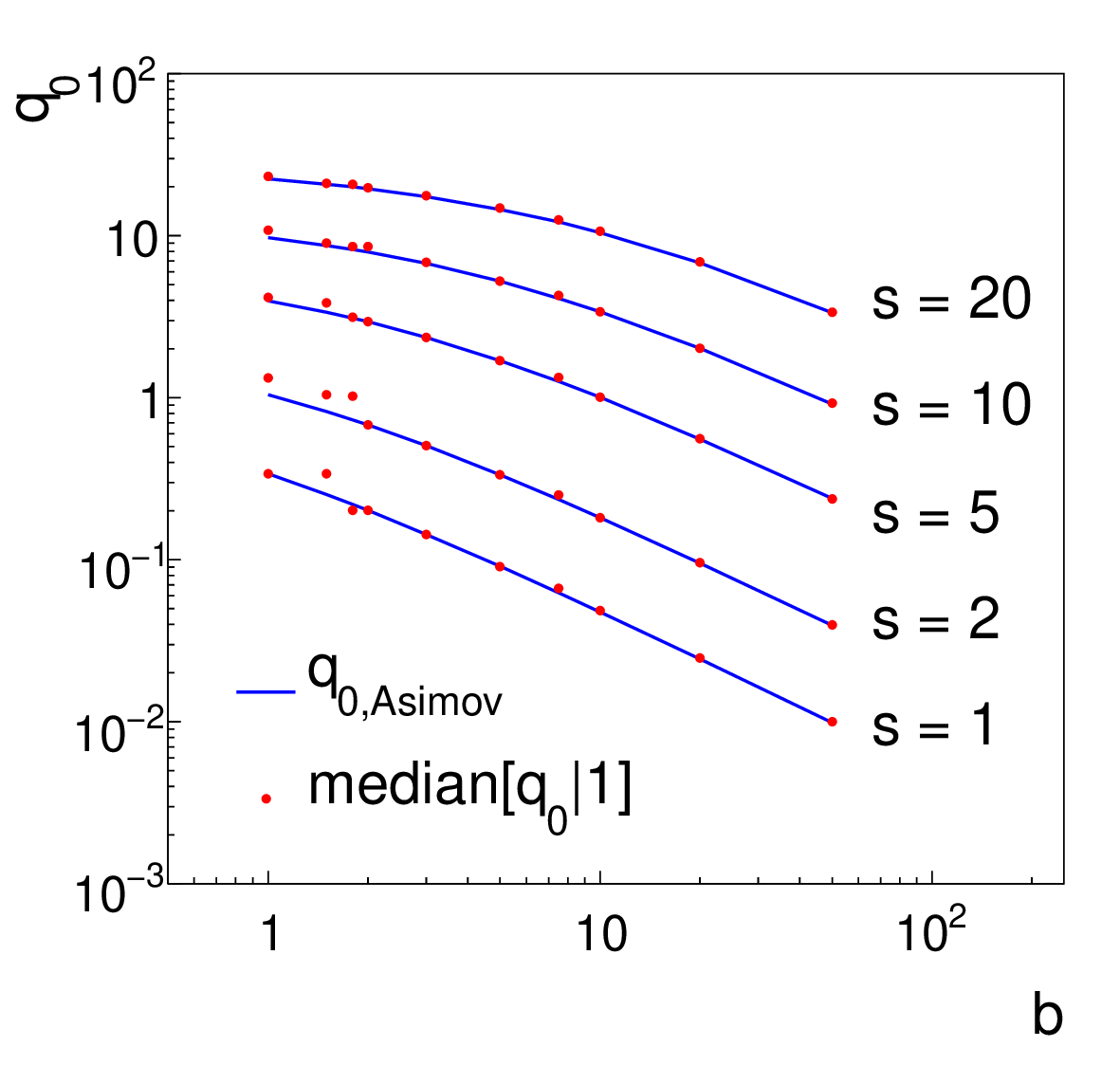}}
\put(0.,6.){(a)}
\put(15,6.){(b)}
\end{picture}
\caption{\small (a) The discovery significance $Z_0$ obtained from Monte
Carlo (points) corresponding to a nominal value $Z_0 = \sqrt{q_0} = 4$
(dashed line) as a function of the expected number of background events
$b$, in the counting analysis with a scale factor $\tau = 1$.
(b) The median of $q_0$ assuming
data distributed according to the nominal signal hypothesis
from Monte Carlo for different values of $s$ and $b$
(points) and the corresponding Asimov values (curves).}
\label{fig:q0test}
\end{figure}
\renewcommand{\baselinestretch}{1}
\small\normalsize

Figure~\ref{fig:q0test}(b) shows the median value of the statistic
$q_0$ assuming data distributed according to the nominal signal
hypothesis from Monte Carlo (points) and the value based on the Asimov
data set as a function of $b$ for different values of $s$, using a
scale factor $\tau = 1$.  One can see that the Asimov data set leads
to an excellent approximation to the median, except at very low $s$
and $b$.

Figure~\ref{fig:q1counting}(a) shows the distribution of the test
statistic $q_1$ for $s = 6$, $b = 9$, $\tau = 1$ for data
corresponding to a strength parameter $\mu^{\prime} = 1$ and also
$\mu^{\prime} = 0$.  The vertical lines indicate the Asimov values of
$q_1$ and $\tilde{q}_1$ assuming a strength parameter $\mu^{\prime} =
0$.  These lines correspond to estimates of the median values of the
test statistics assuming $\mu^{\prime} = 0$.  The areas under the
curves $f(q_1|1)$ and $f(\tilde{q}_1|1)$ to the right of this line
give the median $p$-values.

\setlength{\unitlength}{1.0 cm}
\renewcommand{\baselinestretch}{0.9}
\begin{figure}[htbp]
\begin{picture}(10.0,6.5)
\put(.8,0)
{\includegraphics{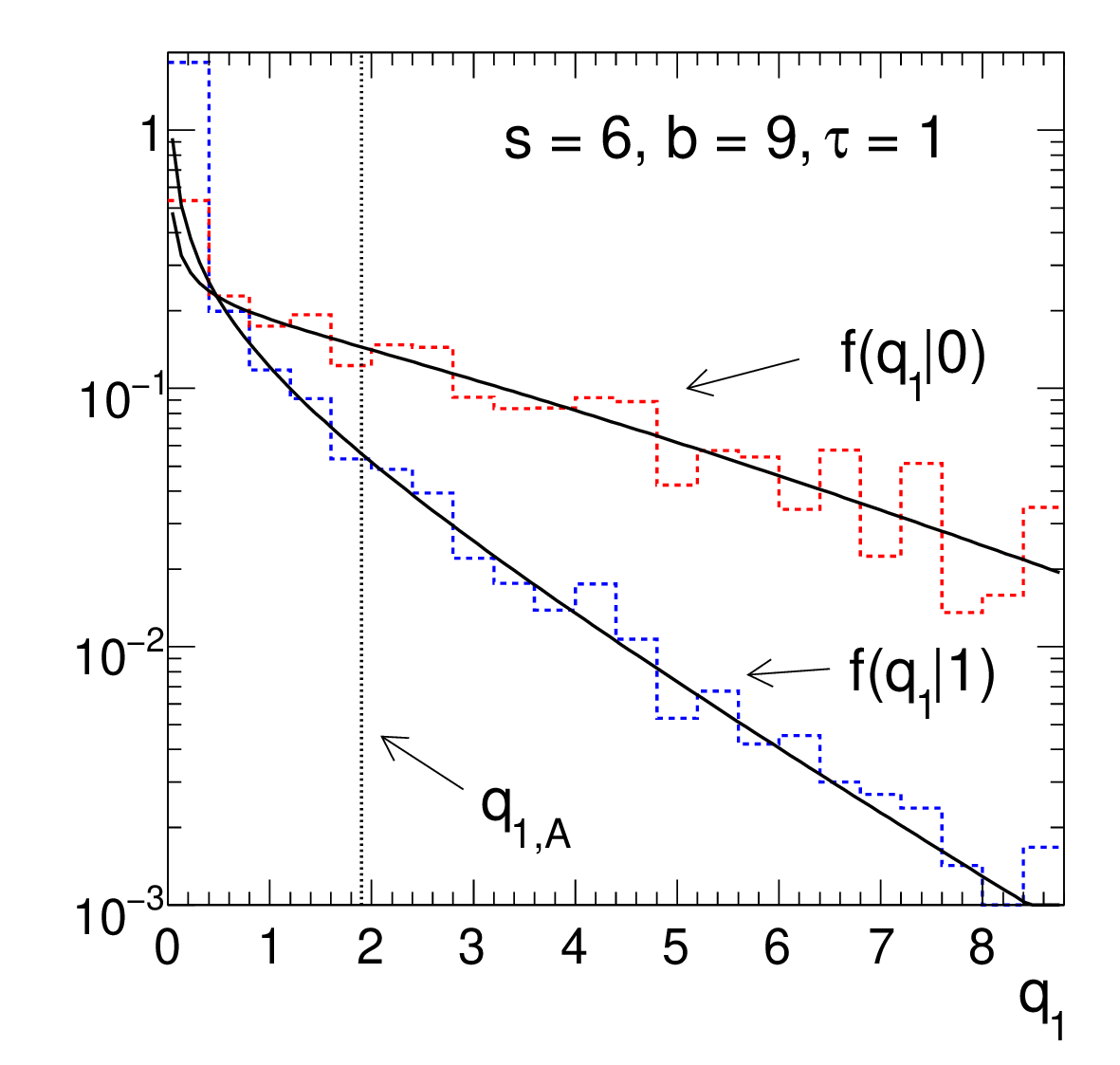}}
\put(8,0)
{\includegraphics{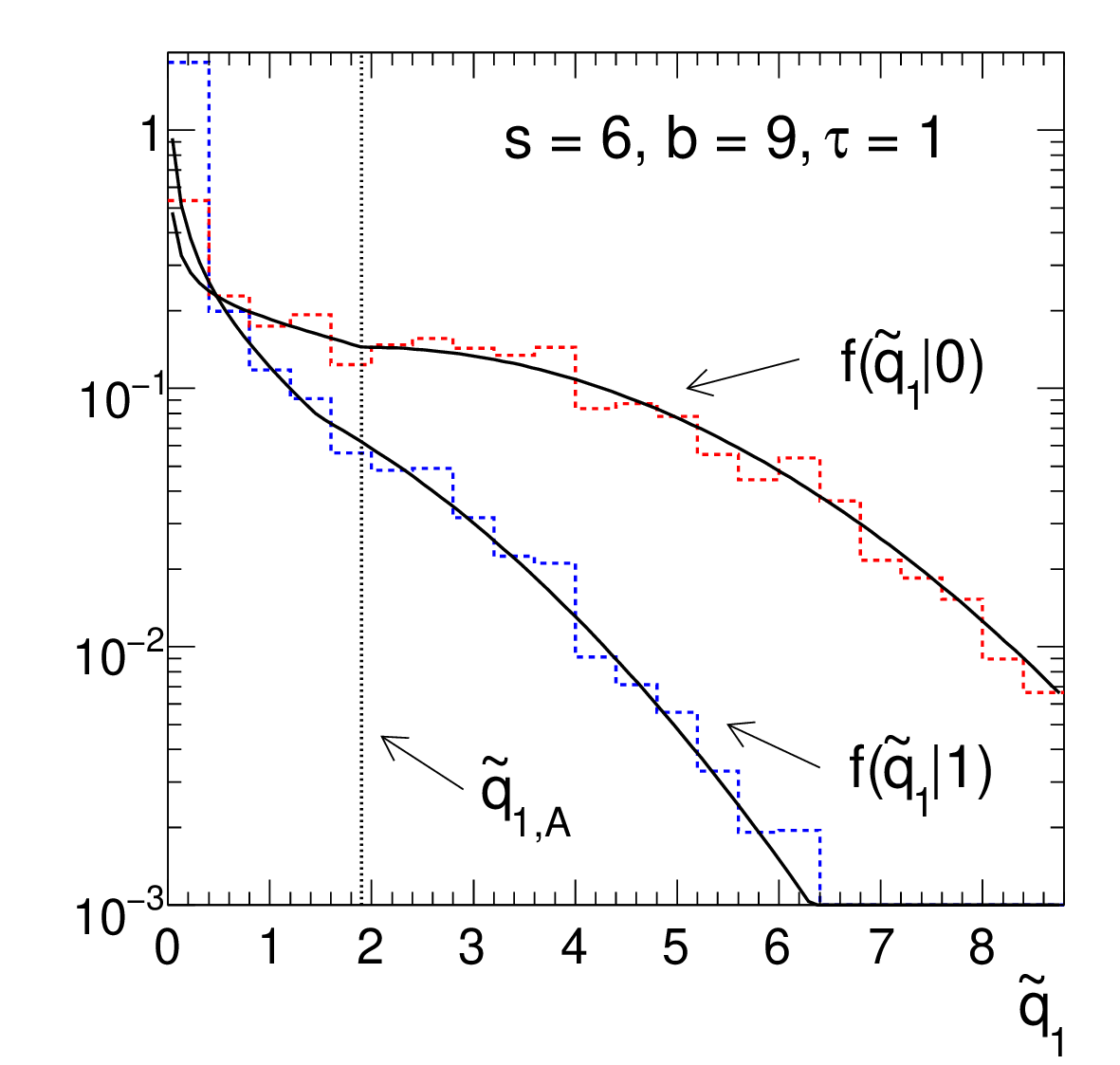}}
\put(0.,6.){(a)}
\put(15,6.){(b)}
\end{picture}
\caption{\small (a) The pdfs $f(q_1|1)$ and $f(q_1|0)$ for the
counting experiment.  The solid curves show the formulae from the
text, and the histograms are from Monte Carlo using $s = 6$, $b = 9$,
$\tau = 1$.  (b) The same set of histograms with the alternative
statistic $\tilde{q}_1$.  The oscillatory structure evident in the
histograms is a consequence of the discreteness of the data.  The
vertical line indicates the Asimov value of the test statistic
corresponding to $\mu^{\prime} = 0$.}
\label{fig:q1counting}
\end{figure}
\renewcommand{\baselinestretch}{1}
\small\normalsize

For the example described above we can also find the distribution of
the statistic $q = - 2 \ln (L_{s+b}/L_{b})$ as defined in
Sec.~\ref{sec:tevatron}.  Figure~\ref{fig:qtevDist} shows the
distributions of $q$ for the hypothesis of $\mu=0$ (background only)
and $\mu=1$ (signal plus background) for the model described above
using $b = 20$, $s = 10$ and $\tau = 1$.  The histograms are from
Monte Carlo, and the solid curves are the predictions of the
asymptotic formulae given in Sec.~\ref{sec:tevatron}.  Also shown are
the $p$-values for the background-only and signal-plus-background
hypotheses corresponding to a possible observed value of the statistic
$q_{\rm obs}$.

\setlength{\unitlength}{1.0 cm}
\renewcommand{\baselinestretch}{0.8}
\begin{figure}[htbp]
\begin{picture}(10.0,6.)
\put(.5,-0.5){\includegraphics{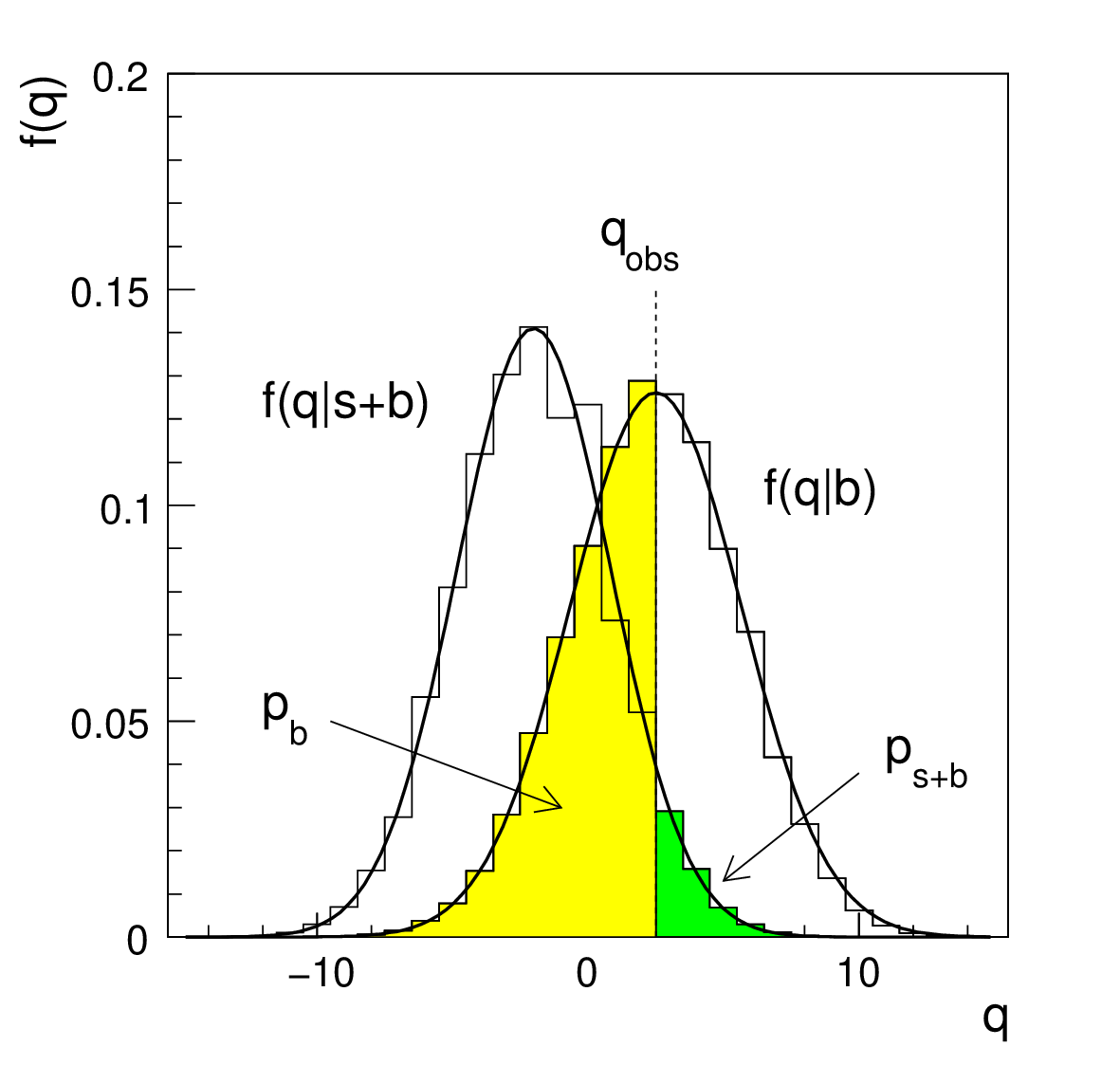}}
\put(9.0,0.){\makebox(6,4)[b]{\begin{minipage}[b]{6cm}
\protect\caption{{\footnotesize The distribution of the
statistic $q = - 2 \ln (L_{s+b}/L_{b})$ under the hypotheses
of $\mu=0$ and $\mu=1$ (see text).  }
\protect\label{fig:qtevDist}}
\end{minipage}}}
\end{picture}
\end{figure}
\renewcommand{\baselinestretch}{1}
\small\normalsize

\subsubsection{Counting experiment with known $b$}
\label{sec:countbknown}

An important special case of the counting experiment above is where
the mean background $b$ is known with negligible uncertainty and can
be treated as a constant.  This would correspond to having a very
large value for the scale factor $\tau$.

If we regard $b$ as known, the data consist only of $n$ and thus the
likelihood function is

\begin{equation}
L(\mu) = \frac{(\mu s + b)^n}{n!} e^{- (\mu s + b)}
\;,
\end{equation}

\noindent The test statistic for discovery $q_0$ can be written

\begin{equation}
\label{eq:q0count}
q_0 = \left\{ \! \! \begin{array}{ll}
    - 2 \ln \frac{L(0)}{L(\hat{\mu})} & \hat{\mu} \ge 0 , \\*[0.3 cm]
               0 & \hat{\mu} < 0 \;,
              \end{array}
       \right.
\end{equation}

\noindent where $\hat{\mu} = n - b$.  For sufficiently large $b$ we
can use the asymptotic formula (\ref{eq:Z0}) for the significance,

\begin{equation}
\label{eq:Z0count}
Z_0 = \sqrt{q_0} = \left\{ \! \! \begin{array}{ll}
               \sqrt{ 2 \left( n \ln \frac{n}{b} + b - n \right) }
                & \hat{\mu} \ge 0 , \\*[0.3 cm]
               0 & \hat{\mu} < 0 .
              \end{array}
       \right.
\end{equation}

To approximate the median significance assuming the nominal
signal hypothesis ($\mu = 1$) we replace $n$ by the Asimov
value $s + b$ to obtain

\begin{equation}
\label{eq:Zmedcount}
\mbox{med}[Z_0 | 1 ] = \sqrt{q_{0,{\rm A}}} =
\sqrt{ 2 \left( (s+b) \ln (1 + s/b) - s \right) } \;.
\end{equation}

\noindent Expanding the logarithm in $s/b$ one finds

\begin{equation}
\label{eq:Zmedcount2}
\mbox{med}[Z_0 | 1 ] = \frac{s}{\sqrt{b}}
\left( 1 + {\cal O}(s/b) \right)\;.
\end{equation}

\noindent Although $Z_0 \approx s/\sqrt{b}$ has been widely used for
cases where $s + b$ is large, one sees here that this final
approximation is strictly valid only for $s \ll b$.

Median values, assuming $\mu=1$, of
$Z_0$ for different values of $s$ and $b$ are shown in
Fig.~\ref{fig:medsig}.  The solid curve shows
Eq.~(\ref{eq:Zmedcount}), the dashed curve gives the approximation
$s/\sqrt{b}$, and the points are the exact median values from Monte
Carlo.  The structure seen in the points is due to the discrete
nature of the data.  One sees that Eq.~(\ref{eq:Zmedcount}) provides a
much better approximation to the true median than does $s/\sqrt{b}$ in
regions where $s/b$ cannot be regarded as small.

\setlength{\unitlength}{1.0 cm}
\renewcommand{\baselinestretch}{0.8}
\begin{figure}[htbp]
\begin{picture}(10.0,6)
\put(.5,-0.5){\includegraphics{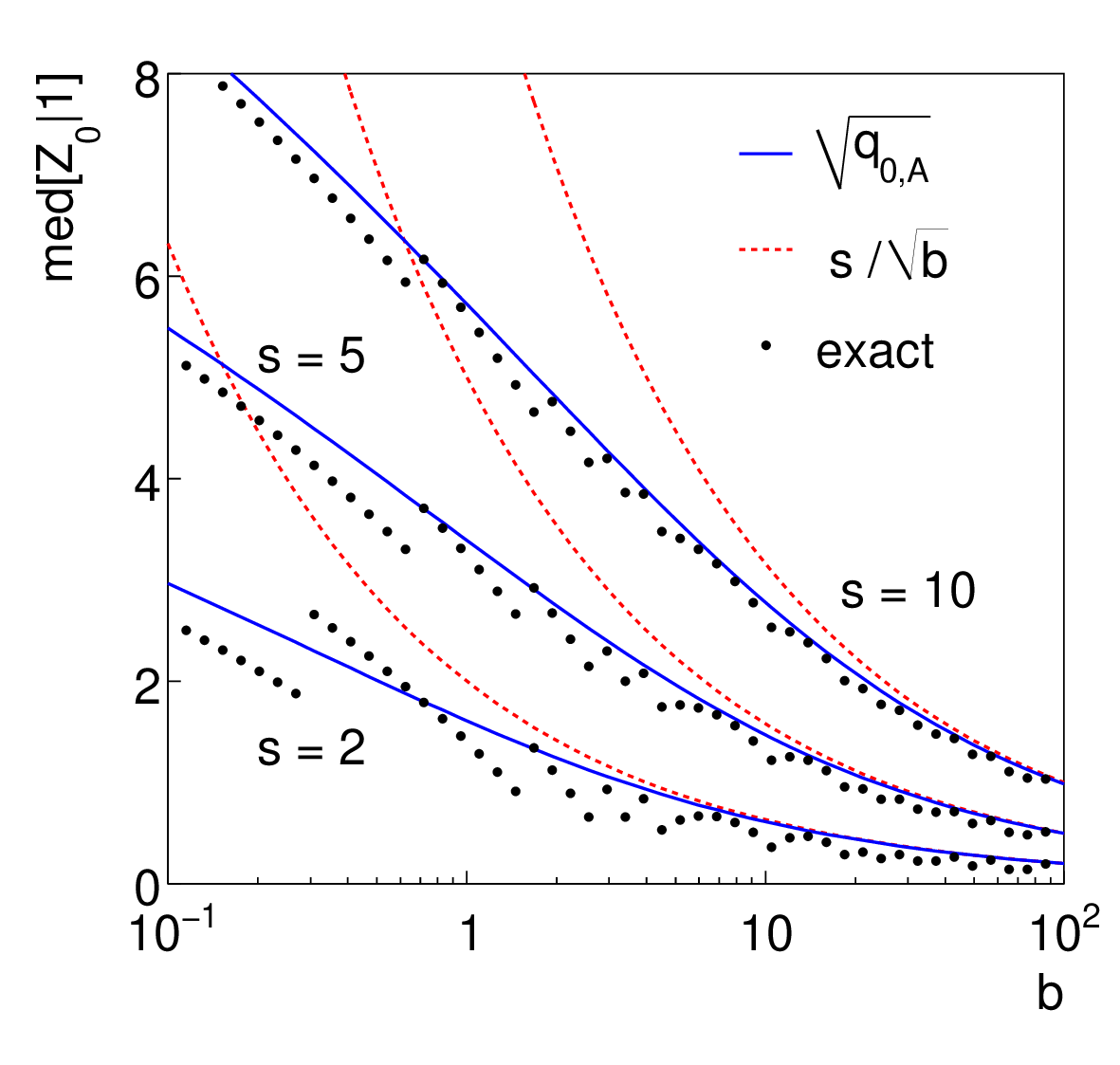}}
\put(9.0,0.){\makebox(6,4)[b]{\begin{minipage}[b]{6cm}
\protect\caption{{\footnotesize The median, assuming $\mu=1$,
of the discovery significance $Z_0$ for different values of
$s$ and $b$ (see text).  }
\protect\label{fig:medsig}}
\end{minipage}}}
\end{picture}
\end{figure}
\renewcommand{\baselinestretch}{1}
\small\normalsize

\subsection{Shape Analysis}
\label{sec:shape}

As a second example we consider the case where one is searching for a
peak in an invariant mass distribution.  The main histogram $\vec{n} =
(n_1, \ldots, n_N)$ for background is shown in
Fig.~\ref{fig:pseudo_exp}, which is here taken to be a Rayleigh
distribution.  The signal is modeled as a Gaussian of known width and
mass (position).  In this example there is no subsidiary histogram
$(m_1, \ldots, m_M)$.

\setlength{\unitlength}{1.0 cm}
\renewcommand{\baselinestretch}{0.8}
\begin{figure}[htbp]
\begin{picture}(10.0,5)
\put(-.5,-4.5){\includegraphics{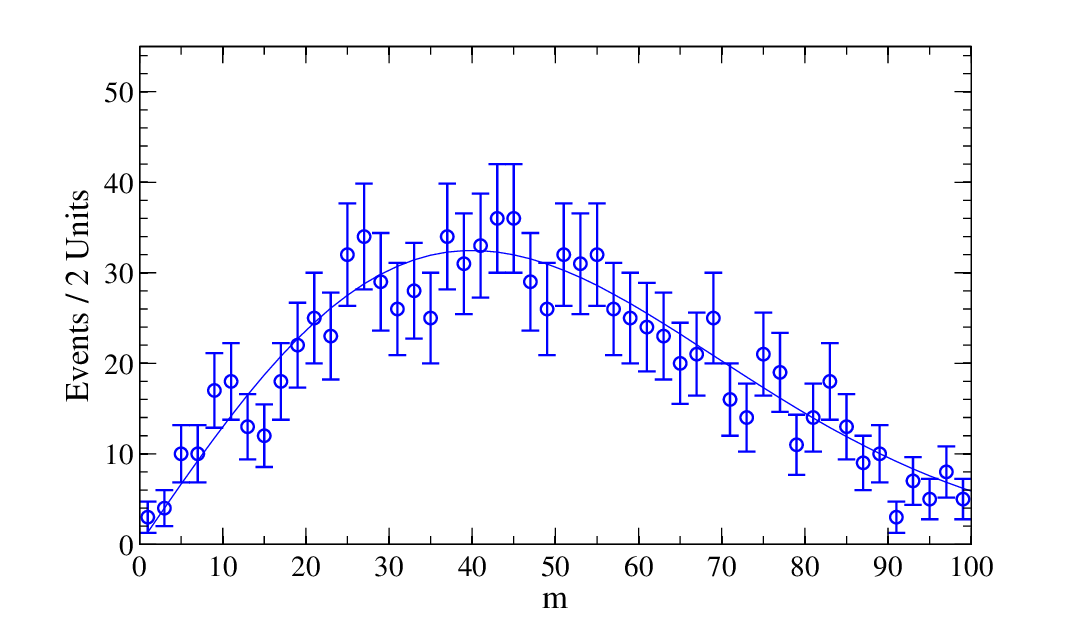}}
\put(10.0,0.){\makebox(5,4)[b]{\begin{minipage}[b]{5cm}
\protect\caption{{\footnotesize The background mass distribution
for the shape analysis (see text).}
\protect\label{fig:pseudo_exp}}
\end{minipage}}}
\end{picture}
\end{figure}
\renewcommand{\baselinestretch}{1}
\small\normalsize

If, as is often the case, the position of the peak is not known a
priori, then one will test all masses in a given range, and appearance
of a signal-like peak anywhere could lead to rejection of the
background-only hypothesis.  In such an analysis, however, the
discovery significance must take into account the fact that
a fluctuation could occur at any mass within the range.  This is often
referred to as the ``look-elsewhere effect'', and is discussed further
in Ref.~\cite{lee}.

In the example presented here, however, we will test all values of the
mass and $\mu$ using the statistic $q_{\mu}$ for purposes of setting
an upper limit on the signal strength.  Here, each hypothesis of mass
and signal strength is in effect tested individually, and thus the
look-elsewhere effect does not come into play.

We assume that the signal and background distributions are known up to
a scale factor.  For the signal, this factor corresponds to the usual
strength parameter $\mu$; for the background, we introduce an
analogous factor $\theta$.  That is, the mean value of the number of
events in the $i$th bin is $E[n_i] = \mu s_i + b_i$, where $\mu$ is
the signal strength parameter and the $s_i$ are taken as known.  We
assume that the background terms $b_i$ can be expressed as $b_i =
\theta f_{{\rm b},i}$, where the probability to find a background event
in bin $i$, $f_{{\rm b},i}$, is known, and $\theta$ is a nuisance
parameter that gives the total expected number of background events.
Therefore the likelihood function can be written

\begin{equation}
\label{eq:likelihoodshape}
L(\mu, \theta) = \prod_{i=1}^N
\frac{(\mu s_i + \theta f_{{\rm b},i})^{n_i}}{n_i!}
e^{-(\mu s_i + \theta f_{{\rm b},i})}
\end{equation}

For a given data set $\vec{n} = (n_1, \ldots, n_N)$ one can evaluate
the likelihood (\ref{eq:likelihoodshape}) and from this determine any
of the test statistics discussed previously.  Here we concentrate on
the statistic $q_{\mu}$ used to set an upper limit on $\mu$, and
compare the distribution $f(q_{\mu} | \mu^{\prime})$ from
Eq.(~\ref{eq:fq0muprimewald}) with histograms generated by Monte
Carlo.  Figure~\ref{fig:qmu_up_dist} shows $f(q_{\mu} | 0)$ (red) and
$f(q_{\mu} | \mu)$ (blue).

\setlength{\unitlength}{1.0 cm}
\renewcommand{\baselinestretch}{0.8}
\begin{figure}[htbp]
\begin{picture}(10.0,5.)
\put(-1,-5){\includegraphics{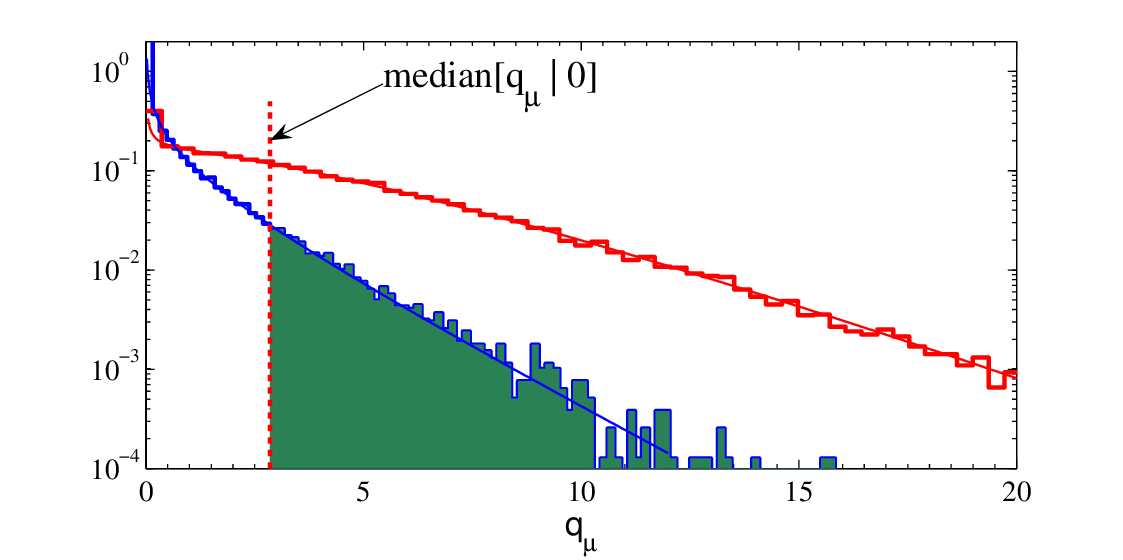}}
\put(10.0,0.){\makebox(5,4)[b]{\begin{minipage}[b]{5cm}
\protect\caption{{\footnotesize The distributions
$f(q_{\mu} | 0)$ (red) and $f(q_{\mu} | \mu )$ (blue) from both
the asymptotic formulae and Monte Carlo histograms (see text).}
\protect\label{fig:qmu_up_dist}}
\end{minipage}}}
\end{picture}
\end{figure}
\renewcommand{\baselinestretch}{1}
\small\normalsize

The vertical line in Fig.~\ref{fig:qmu_up_dist} gives the median value
of $q_{\mu}$ assuming a strength parameter $\mu^{\prime} = 0$.  The
area to the right of this line under the curve of $f(q_{\mu}|\mu)$
gives the $p$-value of the hypothesized $\mu$, as shown shaded in
green.  The upper limit on $\mu$ at a confidence level $\mbox{CL} = 1
- \alpha$ is the value of $\mu$ for which the $p$-value is $p_{\mu} =
\alpha$.  Figure~\ref{fig:qmu_up_dist} shows the distributions for the
value of $\mu$ that gave $p_{\mu} = 0.05$, corresponding to the 95\%
CL upper limit.

In addition to reporting the median limit, one would like to know how
much it would vary for given statistical fluctuations in the data.
This is illustrated in Fig.~\ref{fig:qmu_up+1_dist2}, which shows the
same distributions as in Figure~\ref{fig:qmu_up_dist}, but here the
vertical line indicates the 15.87\% quantile of the distribution
$f(q_{\mu} | 0)$, corresponding to having $\hat{\mu}$ fluctuate
downward one standard deviation below its median.

\setlength{\unitlength}{1.0 cm}
\renewcommand{\baselinestretch}{0.8}
\begin{figure}[htbp]
\begin{picture}(10.0,5.5)
\put(-1,-5){\includegraphics{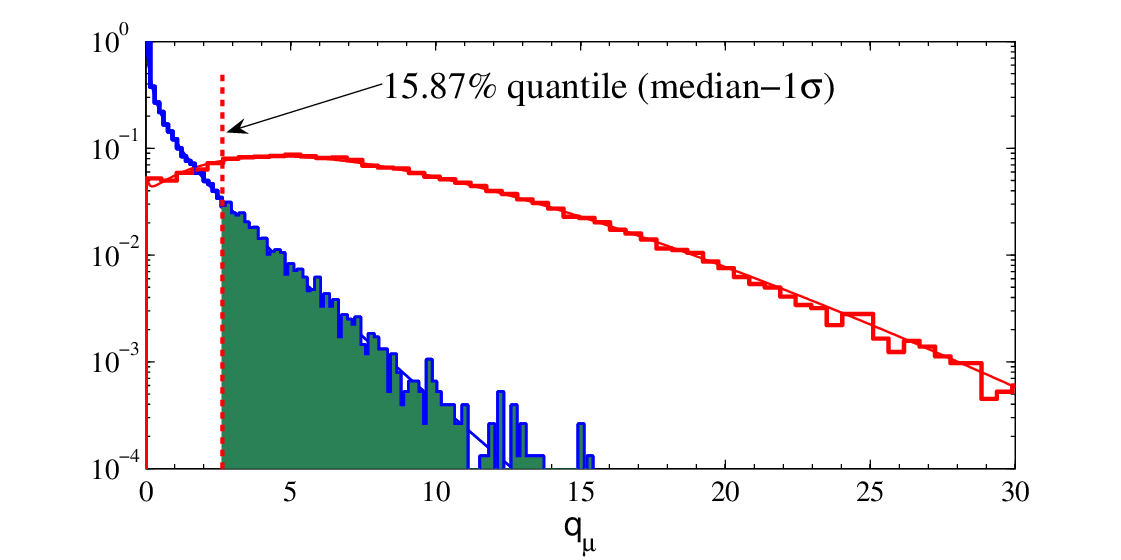}}
\put(10.0,0.){\makebox(5,4)[b]{\begin{minipage}[b]{5cm}
\protect\caption{{\footnotesize The distributions
$f(q_{\mu} | 0)$ (red) and $f(q_{\mu} | \mu )$ (blue)
as in Fig.~\ref{fig:qmu_up_dist} and the 15.87\% quantile
of $f(q_{\mu} | 0)$ (see text).}
\protect\label{fig:qmu_up+1_dist2}}
\end{minipage}}}
\end{picture}
\end{figure}
\renewcommand{\baselinestretch}{1}
\small\normalsize

By simulating the experiment many times with Monte Carlo, we can
obtain a histogram of the upper limits on $\mu$ at 95\% CL, as shown
in Fig.~\ref{fig:mu_up_dist}.  The $\pm 1 \sigma$ (green) and $\pm 2
\sigma$ (yellow) error bands are obtained from the MC experiments.
The vertical lines indicate the error bands as estimated directly
(without Monte Carlo) using Eqs.~(\ref{eq:medmuup}) and
(\ref{eq:muupband}).  As can be seen from the plot, the agreement
between the formulae and MC predictions is excellent.

\setlength{\unitlength}{1.0 cm}
\renewcommand{\baselinestretch}{0.8}
\begin{figure}[htbp]
\begin{picture}(10.0,5.5)
\put(-1,-4.5){\includegraphics{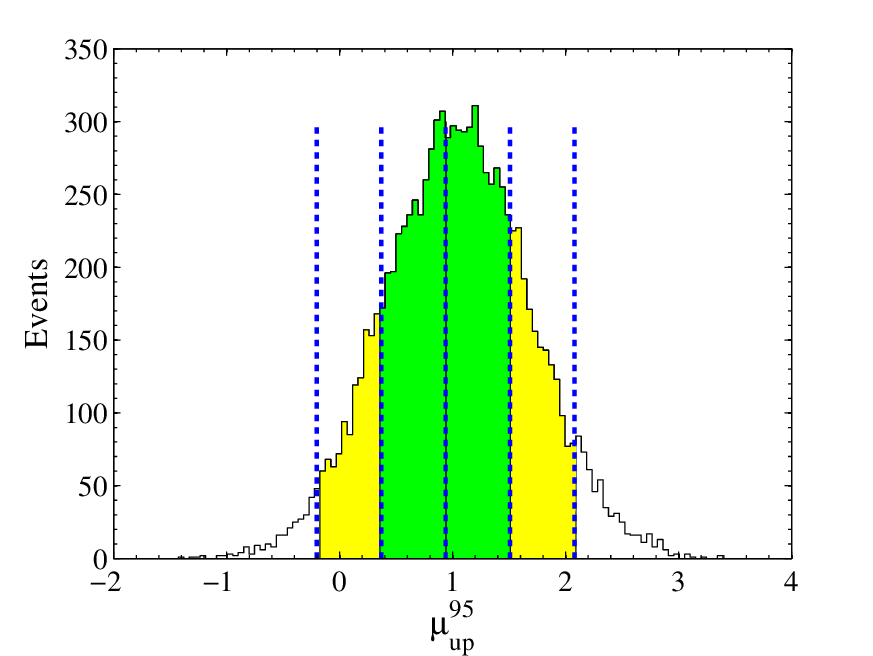}}
\put(10.0,0.){\makebox(5,4)[b]{\begin{minipage}[b]{5cm}
\protect\caption{{\footnotesize Distribution of the upper
limit on $\mu$ at 95\% CL, assuming data corresponding to
the background-only hypothesis (see text).}
\protect\label{fig:mu_up_dist}}
\end{minipage}}}
\end{picture}
\end{figure}
\renewcommand{\baselinestretch}{1}
\small\normalsize

Figures~\ref{fig:qmu_up_dist} through \ref{fig:mu_up_dist} correspond
to finding upper limit on $\mu$ for a specific value of the peak
position (mass).  In a search for a signal of unknown mass, the
procedure would be repeated for all masses (in practice in small
steps).  Figure~\ref{fig:limit_CLsb_modified_1} shows the median upper
limit at 95\% CL as a function of mass.  The median (central blue
line) and error bands ($\pm 1\sigma$ in green, $\pm 2 \sigma$ in
yellow) are obtained using Eqs.~(\ref{eq:medmuup}) and
(\ref{eq:muupband}).  The points and connecting curve correspond to
the upper limit from a single arbitrary Monte Carlo data set,
generated according to the background-only hypothesis.  As can be
seen, most of the plots lie as expected within the $\pm 1 \sigma$
error band.

\setlength{\unitlength}{1.0 cm}
\renewcommand{\baselinestretch}{0.8}
\begin{figure}[htbp]
\begin{picture}(10.0,5.5)
\put(-1,-4.5){\includegraphics{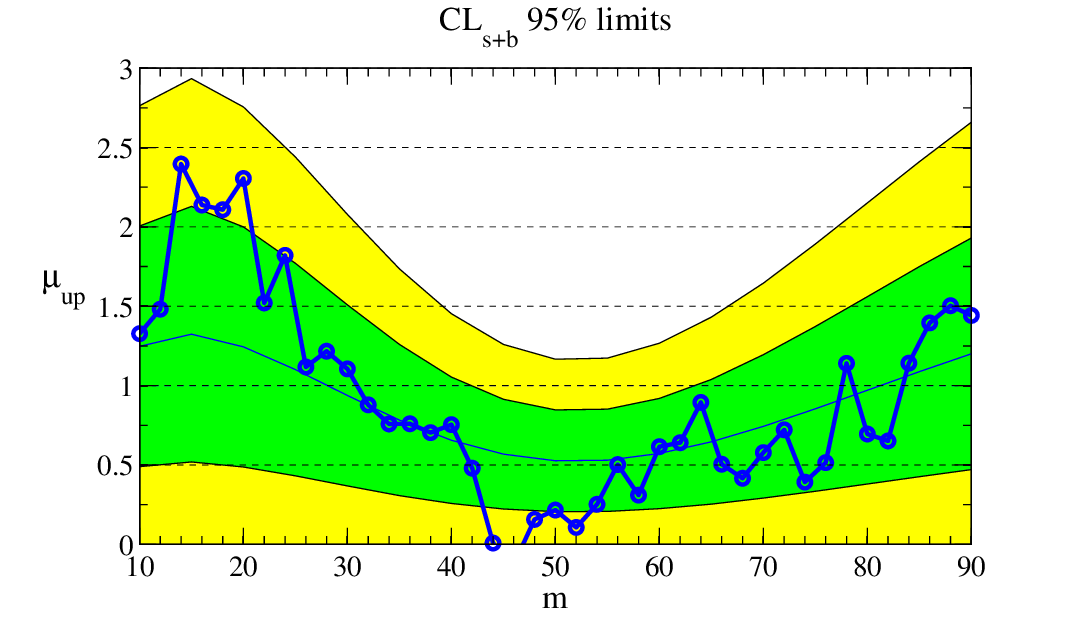}}
\put(10.0,0.){\makebox(5,4)[b]{\begin{minipage}[b]{5cm}
\protect\caption{{\footnotesize The median (central blue
line) and error bands ($\pm 1\sigma$ in green, $\pm 2 \sigma$ in
yellow) for the 95\% CL upper limit on the strength parameter $\mu$
(see text).}
\protect\label{fig:limit_CLsb_modified_1}}
\end{minipage}}}
\end{picture}
\end{figure}
\renewcommand{\baselinestretch}{1}
\small\normalsize

\section{Implementation in RooStats}
\label{sec:roostats}

Many of the results presented above are implemented or are being
implemented in the \mbox{RooStats} framework~\cite{Moneta:2010pm},
which is a C++ class library based on the ROOT~\cite{Brun:1997pa} and
RooFit~\cite{Verkerke:2003ir} packages.  The tools in RooStats can be
used to represent arbitrary probability density functions that inherit
from \texttt{RooAbsPdf}, the abstract interfaces for probability
density functions provided by RooFit.

The framework provides an interface with minimization packages such as
\texttt{Minuit}~\cite{James:1975dr}.  This allows one to obtain the
estimators required in the the profile likelihood ratio: $\hat{\mu}$,
$\hat{\vec{\theta}}$, and $\hat{\hat{\vec{\theta}}}$.  The Asimov
dataset defined in Eq.~(\ref{eq:asimovn}) can be determined for a
probability density function by specifying the \texttt{ExpectedData()}
command argument in a call to the \texttt{generateBinned} method.  The
Asimov data together with the standard \texttt{HESSE} covariance
matrix provided by \texttt{Minuit} makes it is possible to determine
the Fisher information matrix shown in Eq.~(\ref{eq:invcov}), and thus
obtain the related quantities such as the variance of $\hat{\mu}$ and
the noncentrality parameter $\Lambda$, which enter into the formulae
for a number of the distributions of the test statistics presented
above.

The distributions of the various test statistics
and the related formulae for $p$-values, sensitivities and confidence
intervals as given in Sections~\ref{sec:formalism}, \ref{sec:qdist}
and \ref{sec:sensitivity} are being incorporated as well.
%
%
%
RooStats currently includes the test statistics $t_\mu$,
$\tilde{t}_\mu$, $q_0$, and $q$, $q_\mu$, and $\tilde{q}_\mu$ as
concrete implementations of the \texttt{TestStatistic} interface.
Together with the Asimov data, this provides the ability to calculate
the alternative estimate, $\sigma_A$, for the variance of $\hat{\mu}$
shown in Eq.~(\ref{eq:sigma2}).  The noncentral chi-square
distribution is being incorporated into both RooStats and ROOT's
mathematics libraries for more general use.  The various
transformations of the noncentral chi-square used to obtain
Eqs.~(\ref{eq:ftmumPrime}), (\ref{eq:ftildetmmp}),
(\ref{eq:fq0muprimewald}), (\ref{eq:fqmmp}), and (\ref{eq:ftildeqmmp})
are also in development in the form of concrete implementations of the
\texttt{SamplingDistribution} interface.  Together, these new classes
will allow one to reproduce the examples shown in
Section~\ref{sec:examples} and to extend them to an arbitrary model
within the RooStats framework.

\section{Conclusions}
\label{sec:conclusions}

Statistical tests are described for use in planning and carrying out a
search for new phenomena.  The formalism allows for the treatment of
systematic uncertainties through use of the profile likelihood ratio.
Here a systematic uncertainty is included to the extent that the model
includes a sufficient number of nuisance parameters so that for at
least some point in its parameter space it can be regarded as true.

Approximate formulae are given for the distributions of test
statistics used to characterize the level of agreement between the
data and the hypothesis being tested, as well as the related
expressions for $p$-values and significances.  The statistics are
based on the profile likelihood ratio and can be used for a two-sided
test of a strength parameter $\mu$ ($t_{\mu}$), a one-sided test for
discovery ($q_0$), and a one-sided test for finding an upper limit
($q_{\mu}$ and $\tilde{q}_{\mu}$).  The statistic $\tilde{t}_{\mu}$
can be used to obtain a ``unified'' confidence interval, in the sense
that it is one- or two-sided depending on the data outcome.

Formulae are also given that allow one to characterize the sensitivity
of a planned experiment through the median significance of a given
hypothesis under assumption of a different one, e.g., median
significance with which one would reject the background-only
hypothesis under assumption of a certain signal model.  These exploit
the use of an artificial data set, the ``Asimov'' data set, defined so
as to make estimators for all parameters equal to their true values.
Methods for finding the expected statistical variation in the
sensitivity (error bands) are also given.

These tools free one from the need to carry out lengthy Monte Carlo
calculations, which in the case of a discovery at $5 \sigma$
significance could require simulation of around $10^8$ measurements.
They are are particularly useful in cases where one needs to estimate
experimental sensitivities for many points in a multidimensional
parameter space (e.g., for models such as supersymmetry), which would
require generating a large MC sample for each point.

The approximations used are valid in the limit of a large data sample.
Tests with Monte Carlo indicate, however, that the formulae are in
fact reasonably accurate even for fairly small samples, and thus can
have a wide range of practical applicability.  For very small samples
and in cases where high accuracy is crucial, one is always free to
validate the approximations with Monte Carlo.


\section*{Acknowledgements}

The authors would like to thank Louis Fayard, Nancy Andari,
Francesco Polci and Marumi Kado for fruitful discussions.  We
received useful feedback at the Banff International Research
Station, specifically from Richard Lockhart and Earl Lawrence. We
thank Christian Gumpert for spotting a typographical error in the
previous version of the paper. One of us (E.G.) is obliged to the
Benoziyo Center for High Energy Physics, to the the Israeli Science
Foundation(ISF), the Minerva Gesellschaft and the German Israeli
Foundation (GIF) for supporting this work.  K.C. is supported by US
National Science Foundation grant PHY-0854724.  G.C.\ thanks the
U.K.\ Science and Technology Facilities Council as well as the
Einstein Center at the Weizmann Institute of Science, where part of
his work on this paper was done.

\end{document}